\documentstyle[epsfig,12pt]{article}
\newcommand{\fslash}[1]{\mbox{$\!\not\!#1$}}
\newcommand{\Lag}{{\cal L}}

\newcommand{\be}{\begin{equation}}
\newcommand{\ee}{\end{equation}}
\newcommand{\ba}{\begin{eqnarray}}
\newcommand{\ea}{\end{eqnarray}}

\begin{document}

\baselineskip 4 ex

\title{The generalized parton distributions of the nucleon
in the NJL model based on the Faddeev approach
\footnote{Correspondence to: H. Mineo, E-mail: mineo@phys.ntu.edu.tw}}

\author{
 H. Mineo $^{a,b}$, Shin Nan Yang $^a$, Chi-Yee Cheung $^b$, and W. Bentz$^c$
\\
 $^a$ Department of Physics  and \\National Center
 for Theoretical Sciences at Taipei,\\
 National Taiwan University,
Taipei 10617, Taiwan\\
 $^b$ Institute of Physics, Academia Sinica, 
Taipei 11529, Taiwan\\
$^c$ Department of Physics, Tokai University,\\
Hiratsuka-shi, Kanagawa 259-1292, Japan}
\date{\today}
\maketitle

\begin{abstract}

We study the generalized parton distributions, including the 
helicity-flip ones, using  
Nambu-Jona-Lasinio model based on a relativistic Faddeev
approach with `static approximation'. Sum rules relating
the generalized parton distributions to nucleon
electromagnetic form factors are satisfied. Moreover,
quark-antiquark contributions in the region $-\xi<x<\xi$
are non-vanishing. Our results are qualitatively similar to
those calculated with Radyushkin's double distribution
ansatz using forward parton distribution functions
calculated in the NJL model as inputs.

\end{abstract}
\newpage
\section{Introduction}

Together with static properties, electromagnetic form
factors and parton distribution functions are traditionally
the main sources of information on the internal structure
of the nucleon. The electromagnetic form factors of the
nucleon describe the distributions of charge and
magnetization within the nucleon, and they are determined
from the electron-nucleon elastic scattering. Nucleon
parton distribution functions are measured in deep
inelastic scattering of leptons. It is well known that in
the Bjorken limit, the deep inelastic scattering data can
be interpreted in a simple and intuitive picture of
incident letpons scattered by point-like and asymptotically
free partons inside the nucleons.  The parton densities
extracted from these processes encode the distributions of
longitudinal momentum and polarization carried by quarks,
antiquarks, and gluons
within a fast moving nucleon.\\

With the advent of a new generation of high energy high
luminosity lepton accelerators, a wide variety of exclusive
processes in the Bjorken limit become experimentally feasible
\cite{Sabatie02,Burkert03}. Theoretically, it has been
shown that \cite{Radyu96,Ji98,Collins99}, just like deep
inelastic scattering, these exclusive processes are also
factorizable within the framework of perturbative QCD, so
that the hard (short-distance) part is calculable, and the
soft (long-distance) part can be parameterized as universal
generalized parton distributions (GPDs). The GPDs provide
information on parton transverse as well as longitudinal
momentum distributions. Furthermore, apart from the parton
helicities, they also contain information on their orbital
angular momenta.  Hence measurement of GPDs would allow us to
determine the quark orbital angular momentum contribution
to the proton spin, and provide a test of the angular
momentum sum rule for proton \cite{Jaffe90}. In the limit
of vanishing transverse momentum transfer, the GPDs reduce
to the familiar parton distribution functions. Furthermore
their first moments give the nucleon form factors
\cite{Ji97}. So the GPDs provide a connection between
nucleon properties obtained in inclusive (parton
distributions) and elastic (form factors) reactions, giving
us considerable amount of new information on the structure
of the nucleon. Excellent reviews on the GPDs can be found
in Refs. \cite{Ji98a,Goeke01,Diehl03}.

GPDs can be measured in deeply virtual Compton scattering
(DVCS) and in hard exclusive leptoproduction  of mesons.
First measurements of DVCS related to GPSs have been
recently reported in \cite{Hermes01,CLAS01,ZEUS01}, and
other experiments designed to measure GPDs in exclusive
reactions are expected to be carried in the near future
\cite{Sabatie02,Burkert03,EXP1,EXP2}. Hence theoretical
estimates of the GPDs will provide a very useful guide to
future experimental efforts.

Unlike parton distributions, the GPDs in general cannot be
interpreted as particle densities, they are instead
probability amplitudes.  However, like the parton
distribution functions, GPDs reflect the low energy
internal structure of the nucleon, and are at present not directly
calculable from first principle in QCD.  A first attempt to
calculate the first moments of GPDs in quenched lattice QCD
at large pion mass has recently been reported
\cite{Goeckeler04}. However, there is still a considerable large gap
in quark mass to bridge between the state-of-art lattice QCD calculations
and the chiral limit. Other theoretical calculations have
also been performed in various QCD-motivated models of
hadron structures such as MIT bag models
\cite{JI97a,Anikin02}, chiral quark-soliton model \cite{Petrov98},
light-front model \cite{Tiburzi01}, Bethe-Salpeter approach
\cite{Tiburzi02}, and constituent quark models
\cite{Boffi03,Scopetta03}. In this work, we calculate the
nucleon GPDs in the NJL model \cite{NJL}.

One of the most important features of QCD is chiral
symmetry and its spontaneous breaking which dictate  the
hadronic physics at low energy. As an effective quark
theory in low energy region,  NJL model \cite{NJL} is known
to conveniently incorporate these essential aspects of QCD.
Models based on the NJL type of Lagrangians have been very
successful in describing low-energy mesonic physics
\cite{Klevansky92}. Based on relativistic Faddeev equation
the NJL model has also been applied to the baryon systems
\cite{Huang94,Ishii95}.  It has been shown that, using the
quark-diquark approximation, one can explain the nucleon
static properties reasonably well \cite{Asami95,Buck92}. If
one further takes the static quark exchange kernel
approximation, the Faddeev equation can be solved
analytically. The resulting foward parton distribution
functions \cite{Mineo99} successfully reproduce the
qualitative features of the empirical valence quark
distribution \cite{DUR}. Recently, NJL model has been used
to investigate the quark light cone momentum distributions
in nuclear matter and the structure function of a bound
nucleon \cite{Mineo04} as well. In this work, we extend
such a NJL-Faddeev approach to calculate the nucleon GPDs.
Since NJL model is a relativistic field theory, the GPDs so
obtained will automatically satisfy all the general
properties such as the positivity constraints and sum rules
\cite{Diehl03}.

This paper is organized as follows: in
Section 2 we explain the model used in this work. In
Section 3 we outline the calculation of GPDs. Results and
discussions are given in Section 4, and finally a summary
is given in Section 5.

\section{The NJL model for the nucleon}

The SU(2)$_f$ NJL model is characterized by a chirally
symmetric four-fermi contact interaction Lagrangian
$\Lag_I$. With the use of Fierz transformations, the
original NJL interaction Lagrangian $\Lag_I$ can be
rewritten in a form where the interaction strength in any
channel can be read off directly \cite{Ishii95}. In
particular, we are interested in the following channels:
\begin{eqnarray}
\Lag_{I,\pi} &=& G_{\pi} \left[({\bar \psi}\psi)^2 - ({\bar
\psi}
\gamma_5 \mbox{\boldmath $\tau$}  \psi)^2 \right] ,
\label{lpi}\\
{\cal L}_{I,s} &=& G_{s} \left[ {\bar \psi}(\gamma_{5} C ) \tau_2
\beta^A {\bar \psi}^{T}\right]
\left[ \psi^T (C^{-1} \gamma_5 )\tau_2 \beta^{A}\psi \right] ,
\label{ls}
\end{eqnarray}
where  $\beta^A =\sqrt{\frac{3}{2}} \lambda^A$ (A=2,5,7)
are the color $\overline{3}$ matrices, and $C=i\gamma_2
\gamma_0$. $\Lag_{I,\pi}$ represents the interaction in the
$0^+$ and $0^-$ $q\bar q$ channels corresponding to the
sigma meson and the pion, respectively. $\Lag_{I,s}$
describes the $qq$ interaction in the scalar diquark
channel $(J^\pi = 0^+, T=0)$. The interactions (\ref{lpi})
and (\ref{ls}) are invariant under chiral $SU(2)_L\times
SU(2)_R$ transformation. The coupling constants $G_\pi$ and
$G_s$ are related to the ones appearing in the original
$\Lag_I$ by Fierz transformation. Here we shall for
simplicity take ${\cal L}_{I,\pi}$ and ${\cal L}_{I,s}$ as
starting points, and treat $G_{\pi}$ and $G_s$ as free
parameters.

The reduced t-matrices in the pionic and scalar diquark
channels are given by the following expressions
\cite{Mineo99}: \ba \tau_{\pi}(k) =
\frac{-2iG_{\pi}}{1+2G_{\pi}\Pi_{\pi} (k^2 )},\quad \tau_D
(k) = \frac{4iG_s}{1+2G_s\Pi_D  (k^2 )}, \label{tau_D} \ea
with the ``bubble graph'' contribution given by
\be \Pi_{\pi}(k^2
) = \Pi_D (k^2 )=6i\int\frac{d^4 q}{(2\pi)^4 }\mbox{tr}_D
[\gamma_5 S(q)\gamma_5 S(k+q)] ,\label{pis}
\ee
where $S(q)=1/(\fslash{q}-M_Q+i\epsilon)$ is the Feynmann
propagator and $M_Q$ is the constituent quark mass.

In order to simplify the numerical calculations we
approximate $\tau_D(k)$  by
\be \tau_D(k)\rightarrow 4iG_s
-\frac{i g_D^2}{k^2 -M_D^2}, \label{pole}
\ee
in the actual
calculation, where $M_D$ is the diquark bound state mass.
$g_D^2$ is the residue of the pole of $\tau_D (k)$,
\be g_D^2=-2\left(\frac{\partial
\Pi_D(k^2)}{\partial k^2}\right)^{-1}_{k^2 =M_D^2}.
\label{g_D} \ee

In performing the four-momentum loop integral, we have to
introduce a cutoff scheme. In this paper we will adopt the
Pauli-Villars (PV) regularization scheme \cite{PAU} which
preserves the gauge invariance and can be applied to light-cone (LC),
Euclidean, and Minkowsky space integrals. We will follow
\cite{ARR} to determine the subtracted terms in PV
regularization scheme.

The original definition of PV regularization scheme is defined by
the following substitution in every loop integral:
\be
\frac{1}{k_1^2 -M^2}\cdots \frac{1}{k_N^2 -M^2} \rightarrow
\sum_{i=0}^{n}c_i \left\{ \frac{1}{k_1^2 -M^2-\Lambda_i^2}\cdots
\frac{1}{k_N^2 -M^2 -\Lambda_i^2} \right\}, \label{PV} \ee
where $c_0=1$ and $\Lambda_0=0$. For the convergence of the
loop integrals, we need to impose at least 2 conditions
\be
\sum_{i=0}^{n} c_i =0, \,\,\,\, \sum_{i=0}^n c_i
\Lambda_i^2=0. \ee
Thus, we need at least 2 subtractions. In order to reduce
the number of parameters, we choose $n=2$ and take the
limit $\Lambda_1 \rightarrow \Lambda_2 = \Lambda$. The
reduction formula for PV regularization scheme then becomes
\be
\sum_{i=0}^2 c_i f(\Lambda_i^2)=f(0)-f(\Lambda^2)
+\Lambda^2\frac{\partial f(\Lambda^2)}{\partial \Lambda^2}.
\label{PVform} \ee

In Ref. \cite{Mineo99}, the relativistic Faddeev equation
was solved analytically under the static approximation,
namely the momentum dependence of the quark exchange kernel
is neglected, i.e.,
\be {1\over \fslash{p_q}-M_Q+i\epsilon} \rightarrow
{-1\over M_Q}. \ee
The analytical solution for the quark-diquark T-matrix is
given by
\be
T(p)=\frac{3}{M_Q}\frac{1}{1-\frac{3}{M_Q}\Pi_N(p)},
\label{T(p)} \ee
where $\Pi_N(p)$ is the quark-diquark bubble:
\be
\Pi_N(p)=\int \frac{d^4 k}{(2\pi)^4}S(k)\tau_D (p-k).
\label{QDbubble} \ee

The nucleon mass $M_N$ is obtained from the pole of
quark-diquark T-matrix of Eq. (\ref{T(p)}), whose behavior
near the pole is given by
\be T(p)\rightarrow \sum_s \Gamma_N (p,s) \bar{\Gamma}_N (p,s)
/(p^2-M_N^2 +i\epsilon),\ee
where $\bar{\Gamma}_N = \Gamma^\dagger_N \gamma_0$.

Together with Eq. (\ref{T(p)}),
it leads to the following expression for the nucleon vertex
function $\Gamma_N (p,s)$:
\ba \Gamma_N(p,s) &=& \sqrt{Z_N}u_N(p,s),\\
Z_N &=& -\left(\left.\frac{\partial \Pi_N(p)}{\partial \fslash{p}}\right)^{-1}
\right|_{\fslash{p}=M_N}, \label{Z_N} \ea
where $u_N(p,s)$ is the nucleon Dirac spinor with
normalization ${\bar u}_N(p,s)u_N(p,s)=2M_N$. With this
normalization convention, ${\bar
u}_N(p,s)\gamma^{\pm}u_N(p,s)=2p^{\pm}$ and the nucleon vertex
satisfies the relation
\be \frac{-1}{2p_-}{\bar
\Gamma_N}(p,s)\frac{\partial \Pi_N(p)} {\partial
p_+}\Gamma_N(p,s)=1, \label{normalization} \ee
where the LC
variables are defined by $a^{\pm}=a_{\mp}=(a^0\pm
a^3)/\sqrt{2}$, and
$\vec{a}_{\perp,i}=-{\vec{a}_{\perp}}^i$ for $i=1,2$.


\section{GPDs of the nucleon}

It is well known that inclusive deep inelastic scattering
of leptons from nucleon is described by universal parton
distribution functions. Hard exclusive processes measure
another kind of structure functions called generalized
parton distributions (GPDs) of the nucleon; they can be
diagrammatically depicted in Fig. 1. Just like ordinary
parton distributions, the GPDs are also process independent
universal functions.

\begin{figure}[hbtp]
\begin{center}
\epsfig{file=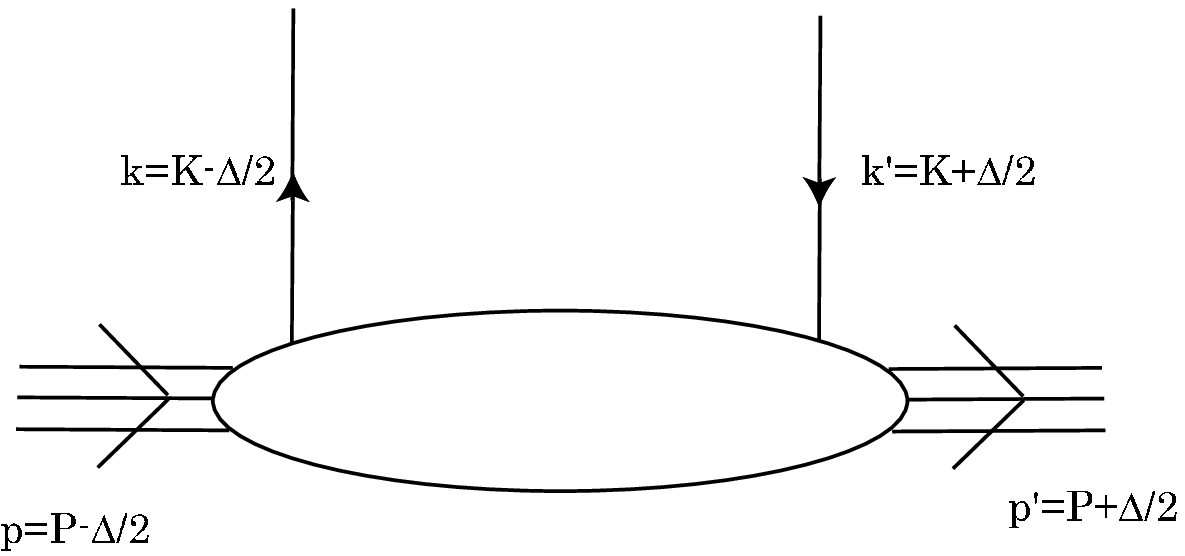,width=7cm} \caption{Soft amplitude for
the GPDs.}
\end{center}
\end{figure}

In hard exclusive processes, a high energy virtual photon
of momentum $q^{\mu}$ is absorbed by a quark in a nucleon,
producing a real photon or a meson and without breaking up
the nucleon \cite{Ji98a,Goeke01,Diehl03}. It is customary
to choose a frame where the averaged nucleon four-momentum
$P=(p+p')/2$ and $q^{\mu}$ are collinear along the z-axis
\cite{Ji98}. Then the GPDs $H(x,\xi,\Delta^2)$ and
$E(x,\xi,\Delta^2)$ are formally given by the leading twist (twist-two)
part of the following amplitude
\ba &&\frac{P_-}{2\pi} \int dy_+ e^{ixP_- y_+}
<p'\lambda'|{\bar \psi}^q(-y/2) \gamma^+ \psi^q
(y/2)|p\lambda>_{y_- =\vec{y}_{\perp} =0}\nonumber\\
&=& {\bar u}_N(p',\lambda')\left[ H^q(x,\xi,\Delta^2)
\gamma^+ +
E^q(x,\xi,\Delta^2)\frac{i\sigma^{+\nu}\Delta_{\nu}}{2M_N}
\right]u_N(p,\lambda) + \cdot\cdot\cdot,
\label{vector current}\nonumber \ea
where $\Delta=p'-p$, superscript $q$ denotes the quark flavor,
$|p\lambda>$ stands for a nucleon state with momentum $p$ and
helicity $\lambda$, and the meaning of $x$ and $\xi$ will be made
clear in the momentum representation below. The ellpsis
$(\cdot\cdot\cdot)$ denotes the higher-twist contributions. In
momentum space the above expression can be written as
\ba &&{\bar u}_N(p',\lambda')\left[
H^q(x,\xi,\Delta^2) \gamma^+ +
E^q(x,\xi,\Delta^2)\frac{i\sigma^{+\nu}\Delta_{\nu}}{2M_N}
\right]u_N(p,\lambda),\nonumber\\
&=& \int \frac{d^4 K}{(2\pi)^4}
\delta(x-K^+/P^+) tr[\gamma^+ \chi_{qN}(p,p',K)]
\ea
where $k=(x+\xi)P^+$ and $k'=(x-\xi)P^+$ are respectively
the initial and final quark momenta, $K=(k+k')/2$, and
$\chi_{qN}(p,p',K)_{ji} =\int d^4 y e^{iK\cdot y}
<p'\lambda'|{\bar \psi}_i(-y/2) \psi_j(y/2) |p\lambda>$ is
the quark-nucleon scattering amplitude. The LC momentum
fraction $x$ and the skewness $\xi$ are given by $x\equiv
K^+/P^+$ and $\xi \equiv -\Delta^+/(2P^+)$, with
\be 0<\xi<\sqrt{\frac{-\Delta^2}{4M_N^2-\Delta^2}}<1.
\label{xi} \ee
Due to the on-shell conditions, $p^2=p'^2=M_N^2$, we also
have:
\be \Delta^2=- \frac{4\xi^2
M_N^2+\vec{\Delta}_{\perp}^2}{1-\xi^2} \ee.

It is then clear that a GPD describes the amplitude of emitting a
parton with momentum fraction $x+\xi$ in a nucleon and reabsorbing
one with momentum fraction $x-\xi$. If $x>\xi$, both the emitted
and absorbed partons are quarks; if $x<-\xi$ then both are
antiquarks.  Finally, if $|x|<\xi$, the two partons involved are a
quark-qntiquark pair.  From this physical picture, it is clear
that, in the forward scattering limit $\xi=0$, GPDs reduce back to
the familiar parton distributions $q(x)$:
\be q(x)=H^q(x,0,0).\ee
Furthermore, by integrating $H^q(x,\xi,\Delta^2)$ over $x$,
we recover the nucleon elastic form factors:
\ba &&\int_{-1}^{1} dx
H^q(x,\xi,\Delta^2)=F_1^q(\Delta^2)\\
&& \int_{-1}^{1} dx E^q(x,\xi,\Delta^2)=F_2^q(\Delta^2).\label{sum rule2}\ea

In the NJL model, the GPDs can be calculated by evaluating
the Feynman diagrams shown in Fig. 2, where the
contributions from the quark and diquark currents,
$J^Q_{\lambda',\lambda}(x,\xi,\Delta^2)$ and
$J^D_{\lambda',\lambda}(x,\xi,\Delta^2)$, are shown
separately. Note that in the NJL model we use here, only
the isoscalar diquark is considered, then it is easy to see
that:
\ba J^{u}_{\lambda',\lambda}(x,\xi,\Delta^2)&=&
J^Q_{\lambda',\lambda} (x,\xi,\Delta^2)
+J^D_{\lambda',\lambda} (x,\xi,\Delta^2),\\
J^{d}_{\lambda',\lambda}(x,\xi,\Delta^2)&=& J^D_{\lambda',\lambda}
(x,\xi,\Delta^2), \label{flavor decomposition} \ea
where superscript $Q(D)$ denotes the quark (diquark) current contribution. We
further write ($X=Q,D$)
\be J^X_{\lambda',\lambda}(x,\xi,\Delta^2) \equiv{\bar
u}_N(p',\lambda')\left[ H^X(x,\xi,\Delta^2) \gamma^+ +
E^X(x,\xi,\Delta^2)\frac{i\sigma^{+\nu}\Delta_{\nu}}{2M_N}
\right]u_N(p,\lambda). \ee

\begin{figure}[hbtp]
\begin{center}
\epsfig{file=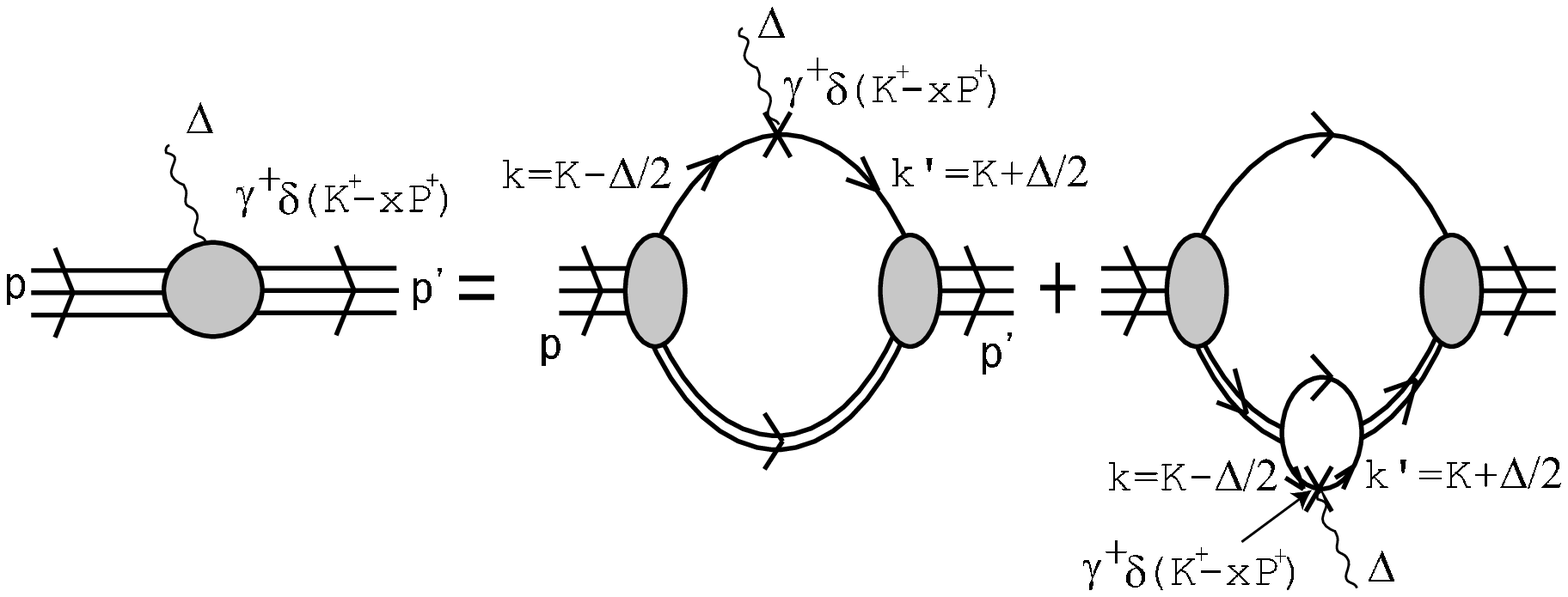,width=13cm}
\caption{ Graphical
representation of the quark GPDs of the nucleon. The single
(double) line denotes the constituent quark propagator
(diquark t-matrix). The operator insertion stands for
$\gamma^+ \delta(K_- -xP_- )(1\pm \tau_z)/2$ for the u(d)
quark. Initial (final) nucleon momentum and helicity are
denoted as $p$ ($p'$) and $\lambda$ ($\lambda'$), and the
four-momentum transfer is given by
$\Delta^{\mu}=p'^{\mu}-p^{\mu}$. }
\end{center}
\end{figure}

Using the table of matrix elements listed in Appendix A, we can
separate the left hand side of Eq. (\ref{vector current}) into
helicity conserving and helicity flipping contributions:
\ba &&J^X_{\lambda',\lambda}(x,\xi,\Delta^2) =
\frac{P^+}{M_N}{\bar u}_N(p',\lambda')u_N(p,\lambda)
\nonumber\\
&\times &\left[ \delta_{\lambda',\lambda}\left(
(1-\xi^2)H^X(x,\xi,\Delta^2)-\xi^2 E^X(x,\xi,\Delta^2) \right)
-\delta_{\lambda',-\lambda}E^X(x,\xi,\Delta^2)
\right]\nonumber\\
\label{helicity relation} \ea

In the following we shall only give an outline of the
calculations, and leave the details to Appendices B and C.

Using simple Feynmann rules, we can directly read off the
quark current contribution $J^Q_{
\lambda',\lambda}(x,\xi,\Delta^2)$ from Fig. 2,
\be J^Q_{\lambda',\lambda} (x,\xi, \Delta^2) = -Z_N{\bar
u}_N(p',\lambda') \int \frac{d^4 K}{(2\pi)^4} \delta \left(
x-\frac{K^+}{P^+} \right) S(k')\gamma^+ S(k)\tau_D (p-k)
u_N(p,\lambda)\\, \label{QC} \ee
where $\tau_D$ is the reduced t-matrix of the diquark.
$\tau_D$ can be decomposed into two terms:
\be \tau_D=\tau_D^C+\tau_D^P \ee
where $\tau_D^C$ and $\tau_D^P$ are respectively the
"contact" and "pole" contributions, as given in
Eq. (\ref{pole})  Accordingly, the quark
current contribution $J^Q$ can also be separated into two
terms:
\be J^Q_{\lambda',\lambda} (x,\xi, \Delta^2)= \theta(-\xi<x<\xi)
J^{Q,C}_{\lambda',\lambda} (x,\xi, \Delta^2) +
\theta(-\xi<x<1)J^{Q,P}_{\lambda',\lambda} (x,\xi, \Delta^2),
\label{J^Q} \ee
where we see that $J^{Q,C}$ contributes only in the region
$-\xi<x<\xi$, while $J^{Q,P}$ only in $-\xi<x<1$.   Thus we see
that, unlike other calculations using non-relativistic quark
models, the field theoretic NJL model gives a non-zero
contribution in the $-\xi<x<1$ region.

Our final expressions for the $H_Q$ and $E_Q$ are given in Eqs.
(\ref{H_Q}) and (\ref{E_Q}).

Similarly, we can also write down the diquark current
contribution,
\ba &&J^D_{\lambda',\lambda}(x,\xi,\Delta^2) =-Z_N{\bar
u}(p',\lambda')\int\frac{d^4 T}{(2\pi)^4}
iS(P-T)\tau_D(t')\tau_D(t)\nonumber\\
&\times& i \int\frac{d^4 K}{(2\pi)^4} tr\left[\gamma^5
C\tau_2 \beta_A S(k')\gamma^+ S(k)C^{-1} \gamma^5 \tau_2
\beta_A S(t-k)^T \right] \delta
\left(x-\frac{K_-}{P_-}\right)u(p,\lambda),\nonumber\\
\label{Diquark current} \ea
where $t=T-\Delta/2, t'=T+\Delta/2$ are the diquark momenta.

We define two additional LC momentum fractions $y, z$,
\be y\equiv K^+/T^+,~~ z\equiv T^+/P^+, \ee
so that
\be \int dy \int dz \delta \left(y-\frac{K^+}{T^+}\right)
\delta \left(z-\frac{T^+}{P^+}\right)=1.
\label{identity}\ee
Inserting Eq. (\ref{identity}) into Eq. (\ref{Diquark current}),
we can rewrite the diquark current contribution in a convolution
form:
\ba
&&J^D_{\lambda',\lambda}(x,\xi,\Delta^2) = \int \frac{d^4
T}{(2\pi)^4}\int dy \int dz \delta (x-yz) {\cal
F}^{D/N}_{\lambda',\lambda}
 (z,\xi,T,\Delta)\nonumber\\
&\times&  \left(F^D_s (y,\zeta,T,\Delta) T^+ +
 F^D_a (y,\zeta,T,\Delta)\Delta^+ \right),
\label{J^D}
\ea
with
\ba
&&\left(F^D_s (y,\zeta,T,\Delta) T^+ + F^D_a
(y,\zeta,T,\Delta)\Delta^+ \right)\nonumber\\
&=&
ig_D^2 \int\frac{d^4 K}{(2\pi)^4}
\delta \left( y-\frac{K^+}{T^+} \right)
tr\left[\gamma^5 C\tau_2 \beta_A S(k')\gamma^+ S(k)C^{-1}
\gamma^5 \tau_2 \beta_A S(t-k)^T \right],\nonumber\\
\label{F_D}
\ea
\be
{\cal F}^{D/N}_{\lambda',\lambda}(z,\xi,T,\Delta) =
-ig_D^{-2} Z_N {\bar u}(p',\lambda')
\delta \left(z-\frac{T^+}{P^+}\right)
S(P-T)\tau_D(t')\tau_D(t)u(p,\lambda),
\label{F_D/N}
\ee
where $g_D$ is given in Eq. (\ref{g_D}), and $\zeta$ is the
skewness defined by the relation $\Delta^+=-2\zeta T^+$.

From the Ward identity for the diquark-diquark-photon vertex
$\Gamma_D^{\mu}(T,\Delta)$,
\be\Delta_{\mu}\tau_D(t')\Gamma_D^{\mu}(T,\Delta)
\tau_D(t)=+2i[\tau_D(t')-\tau_D(t)],\ee
we obtain
\be \tau_D(t')\tau_D(t)= \frac{2i[\tau_D(t')-\tau_D(t)]} {
\Delta_{\mu} \Gamma_ D^{\mu}(T,\Delta) }.
\label{Ward2}\ee
In order to reduce the complexity of the calculation, we introduce
the 'on-shell diquark approximation', i.e., $t^2=t'^2=M_D^2$. Then
the vertex $\Gamma_D^{\mu}$ can be expressed in terms of a single
form factor $G_s^{D}$
\be \Gamma_D^{\mu} (T,\Delta)\simeq G_s^{D}
(\Delta^2)T^{\mu}.\label{onshell}\ee
(see Appendix C for details).

Using Eqs. (\ref{pole}) and (\ref{onshell}), we can rewrite
Eq. (\ref{Ward2}) as
\be \tau_D(t')\tau_D(t) \rightarrow
\frac{-g_D^2}{G_s^{D}(\Delta^2) (t'^2 -M_D^2)(t^2-M_D^2)}.
\ee
Substituting this result into Eq. (\ref{F_D/N}), we finally
arrive at
\be {\cal F}^{D/N}_{\lambda',\lambda}(z,\xi,T,\Delta) =
iZ_N {\bar u}(p',\lambda') \frac{\delta (z-T^+/P^+)S(P-T)}
{G_s^{D}(\Delta^2)(t'^2-M_D^2)(t^2-M_D^2)}u(p,\lambda).
\label{F_{D/N}} \ee
The final results for $H^D(x,\xi,\Delta^2)$ and
$E^D(x,\xi,\Delta^2)$ are given in Eq. (\ref{H_D}). In Appendix C,
apart from $\Delta^{\mu}$, we have introduced another momentum
transfer variable $\Delta_D^{\mu}$ inside the convolution
integral. In a complete evaluation, we should have
$\Delta^{\mu}=\Delta_D^{\mu}$.  However the on-shell diquark
approximation gives raise to an ambiguity. In \cite{Scopetta03},
it is assumed that $\Delta^2=\Delta_D^2$ and
$\Delta^+=\Delta_D^+$, then $\zeta=\xi/z$. This form is adopted
for small $\xi$ and $\vec{\Delta}_{\perp}^2<< M_N^2$. However this
choice of $\Delta_D^{\mu}$ is not satisfactory in our case, since
for $\Delta^+=\Delta_D^+$ implies that the GPDs are non-vanishing
only in the region $\xi<z$, and it follows from Eq. (\ref{sum
rule}) that the sum rule is explicitly broken.  In order to
preserve the sum rule relation, which we believe is important, we
let $\vec{\Delta}_{\perp}^2=\vec{\Delta}_{D\perp}^2$.  Then
$\zeta$ is fixed by the relation $\Delta^2=\Delta_D^2$ because
\ba &&\Delta^2=-\frac{4\xi^2 M_N^2
+\vec{\Delta}_{\perp}^2}{1-\xi^2}\label{Delta and xi},\\
&&\Delta_D^2=-\frac{4\zeta^2 M_D^2
+\vec{\Delta}_{D\perp}^2}{1-\zeta^2}. \ea

Finally we also include the photon-quark vertex correction for the
GPDs and electromagnetic form factors which arises from the
structure of the constituent quark. Specifically, we sum the
series of ring diagrams as shown in Fig. 3. In the spirit of
vector dominance, we demand that the resultant photon vertex
possesses a pole at $\Delta^2=M_{\omega}^2 ~(\Delta^2=M_{\rho}^2)$
in the isoscalar (isovector) channel . This effectively replaces
the bare vertex $\tau^a \gamma^{\mu}$ by
\be \tau^a \gamma^{\mu}\rightarrow \frac{\tau^a \gamma^{\mu}}
{1+2G_a \Delta^2 \Pi_{V,T}(\Delta^2)}, \ee
where $a=0(i)$ corresponds to the isoscalar
(isovector) part $(\tau^0=1)$, and the corresponding
coupling constants are $G_{\omega}$ (for $a=0$) and
$G_\rho$ (for $a=i$). The definition of
$\Pi_{V,T}(\Delta^2)$ and the details of the calculation can be
found in appendix D.

\begin{figure}[hbtp]
\begin{center}
\epsfig{file=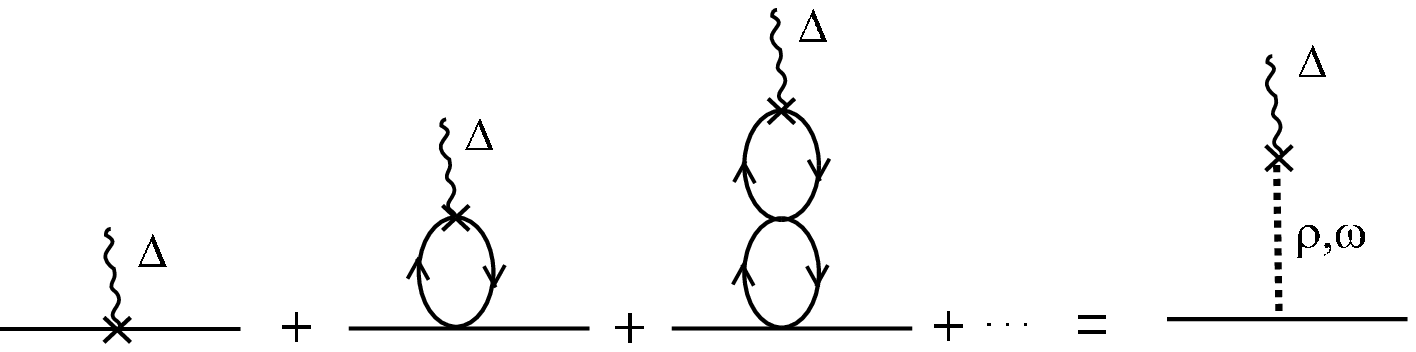,width=13cm} \caption{The vector meson dominance
corrections to the $\gamma qq$ vertex. The dotted line  represents
the vector mesons $\omega$ and $\rho$.}
\end{center}
\end{figure}

As we shall see in the next section, these vertex
corrections significantly improve the momentum dependence
of the electromagnetic form factors calculated in the NJL
model.

\section{Results and discussions}

In this section, we present the results of our calculation.
We shall first explain the choice of parameters in our
model. Subsequently, numerical results are presented and
compared with those obtained from other works.

In the NJL model we have adopted here, the constituent
quark mass is taken to be $M_Q=400$ MeV, which is within
the range of values used in other works \cite{CHR96}. Using
this constituent quark mass, together with the pion mass
$m_{\pi}=140$ MeV and the pion decay constant $f_{\pi}=93$
MeV , we can determine the Pauli-Villars cutoff parameter
$\Lambda=739$ MeV, the coupling constant $G_{\pi}=10.42$
GeV$^{-2}$, and the current quark mass $m_q=9$ MeV.
Furthermore, we set $G_s=0.65G_{\pi}$ so that the
solution of the Faddeev equation reproduces the
experimental nucleon mass $M_N=940$ MeV, the scalar diquark
mass is then fixed to be $M_D=590$ MeV.  Finally, the
coupling constants $G_{\omega}=7.34$ GeV$^{-2}$ and
$G_{\rho}=8.38$ GeV$^{-2}$ are determined from the poles of
Eq. (\ref{vertex corrections}), so that the physical vector
meson masses $m_{\omega}=783$ MeV and $m_{\rho}=770$ MeV
are reproduced.

The nucleon electromagnetic form factors $F_{1}^{p} (\Delta^2)$
and $F_{1}^{n} (\Delta^2)$ calculated in the NJL model are
depicted in Figs. 4(a) and 4(b), respectively, where results with
and without corrections to the photon-quark electromagnetic vertex
(see Fig. 3) are shown separately. For comparison, we have also
plotted the results corresponding to the familiar "dipole fit" to
the experimental data of $G_E$'s and $G_M$'s by dashed lines:
\ba G_E^p,G_M^{p,n}&\propto & (1-\Delta^2/\Delta_0^2)^{-2}\cr
 G_E^n &=& 0,\ea
with $\Delta_0^2=0.71$ GeV$^2$. We see that the effect of the vertex corrections is rather
sizable. In the proton case, where the data are much more
precise, inclusion of the vertex correction significantly
improves the agreement with experimental data, which is
very well parameterized by the ``dipole fit". Nevertheless
discrepancies still exist for $-\Delta^2>0.5$ GeV$^2$. Note
that for $\vec{\Delta}_{\perp}=\vec{0}_{\perp}$,
$-\Delta^2=0.5$ GeV$^2$ corresponds to $\xi\simeq 0.35$,
and in this work we are only concerned with small $\xi$
($\le 0.3$). In the case of the neutron, the effect of the
vertex correction is small. Compared with the ``dipole
fit", the NJL model result for $F_{1}^{n} (\Delta^2)$ is
similar in magnitude, but opposite in sign. Unfortunately
the available data in this case are scattered with
large error bars, so that it is not possible to
determine which curve fits better.

Similarly, the nucleon  form factors $F_2^{p,n}
(\Delta^2)$ are plotted in Figs. 5(a) and (b). We see that
the NJL model results are significantly different from the
dipole fits in the low momentum region $-\Delta^2<1$
GeV$^2$. As a result, the calculated nucleon magnetic
moments (in units of nuclear magneton)
\be \mu^{NJL}_{p}=1.75,~~~  \mu^{NJL}_{n}=-0.82, \ee
are much smaller  than the experimental values
\be \mu^{exp}_{p}=2.79,~~~  \mu^{exp}_{n}=-1.91. \ee
This will affect the reliability of the $E^q(x,\xi,\Delta^2)$
calculated in this model (see discussions below). It is known
that further inclusion of the axial vector diquark channel and
the pion cloud are important to improve the results for the
magnetic moment \cite{Mineo99}. However, these effects are outside the 
scope of our present investigation of GPDs.

\begin{figure}[hbtp]
\begin{center}
\epsfig{file=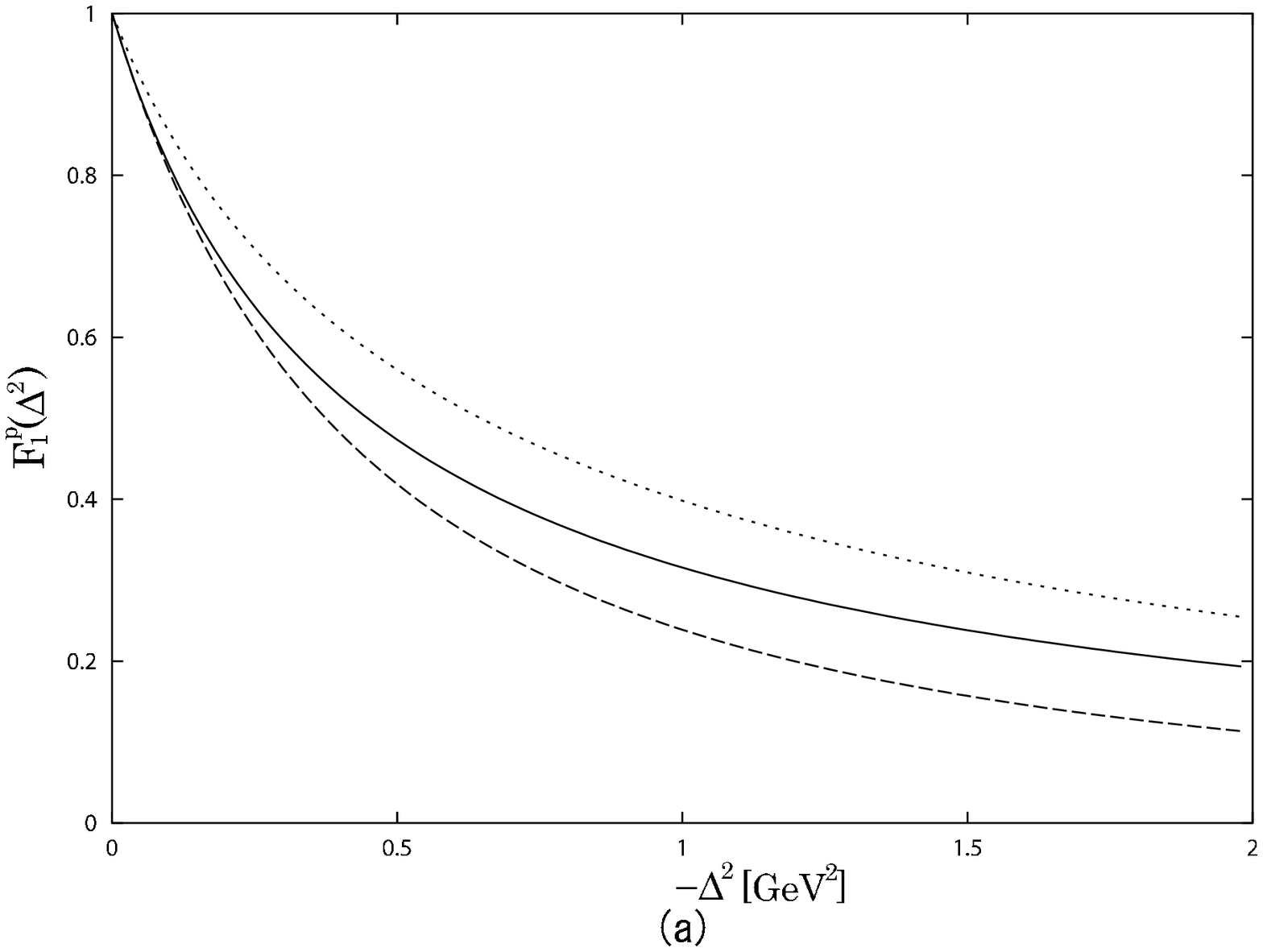,width=9cm}
\epsfig{file=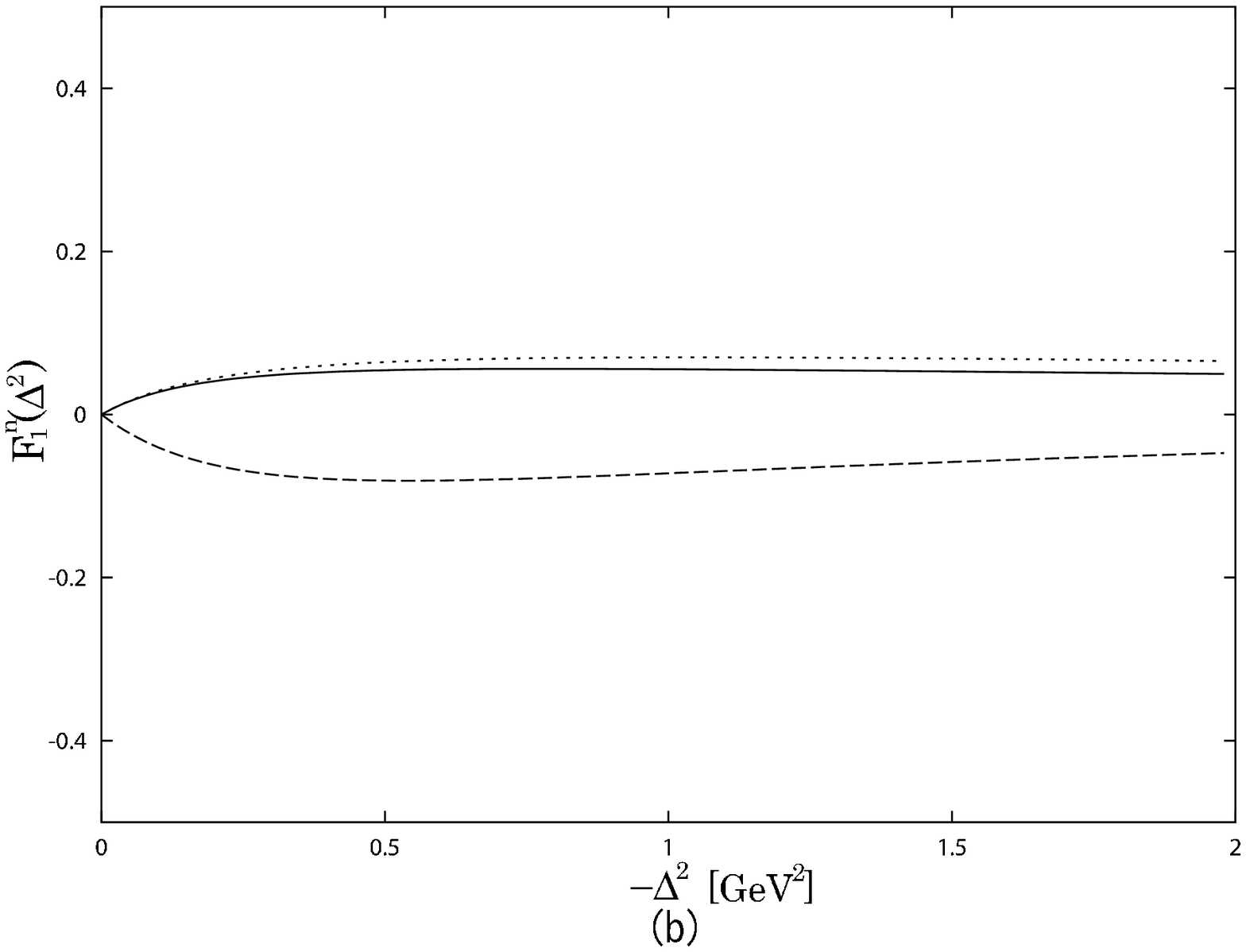,width=9cm}
\end{center}
\caption{(a) Proton form factor $F_1^p(\Delta^2)$.
The dotted and solid lines are calculated in the NJL model without 
and without vertex corrections, respectively,
and the dashed line is the dipole fit to the experimental
data.  (b) Neutron form factor $F_1^n(\Delta^2)$ in the same notation as
(a).}
\end{figure}

\begin{figure}
\begin{center}
\epsfig{file=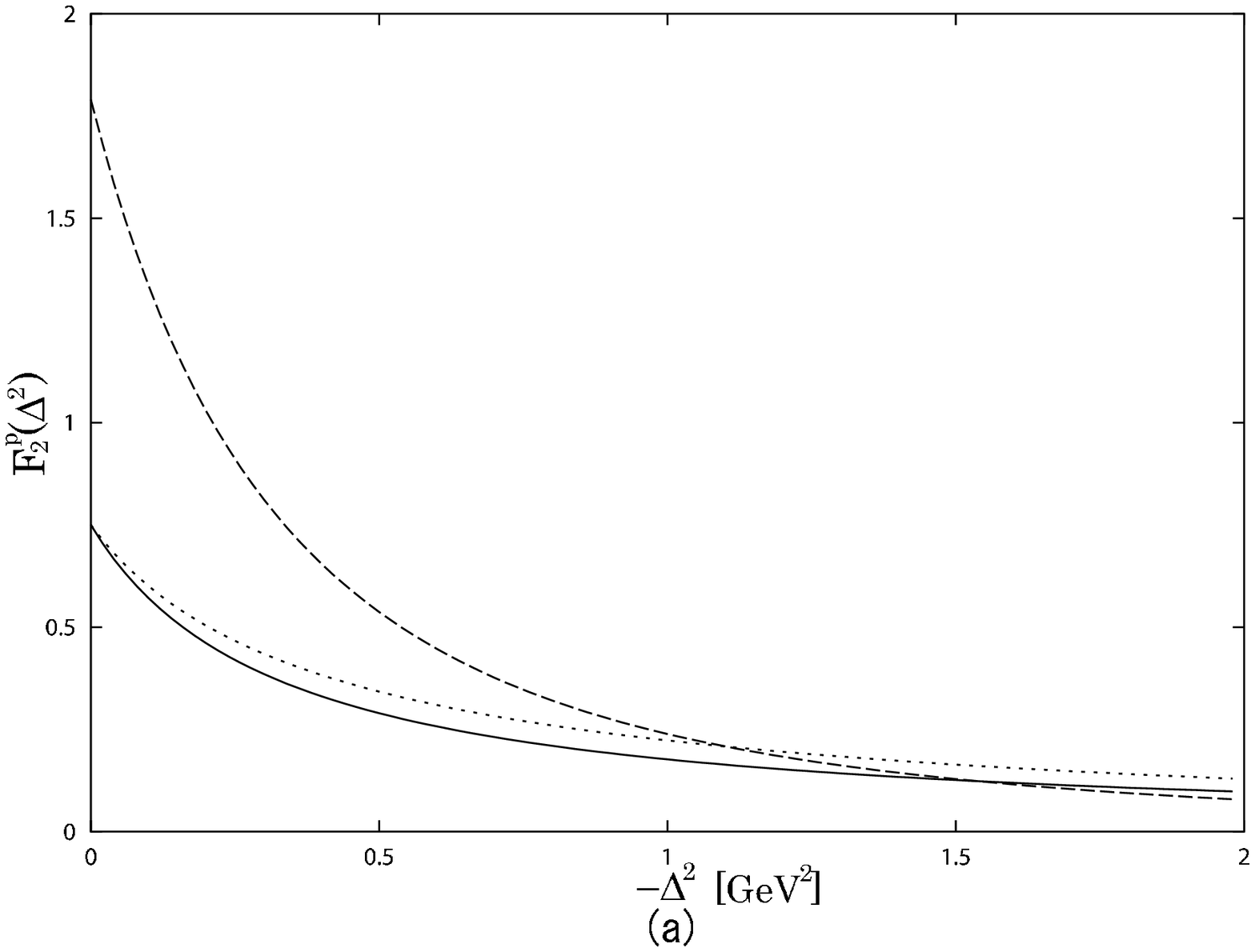,width=9cm}
\epsfig{file=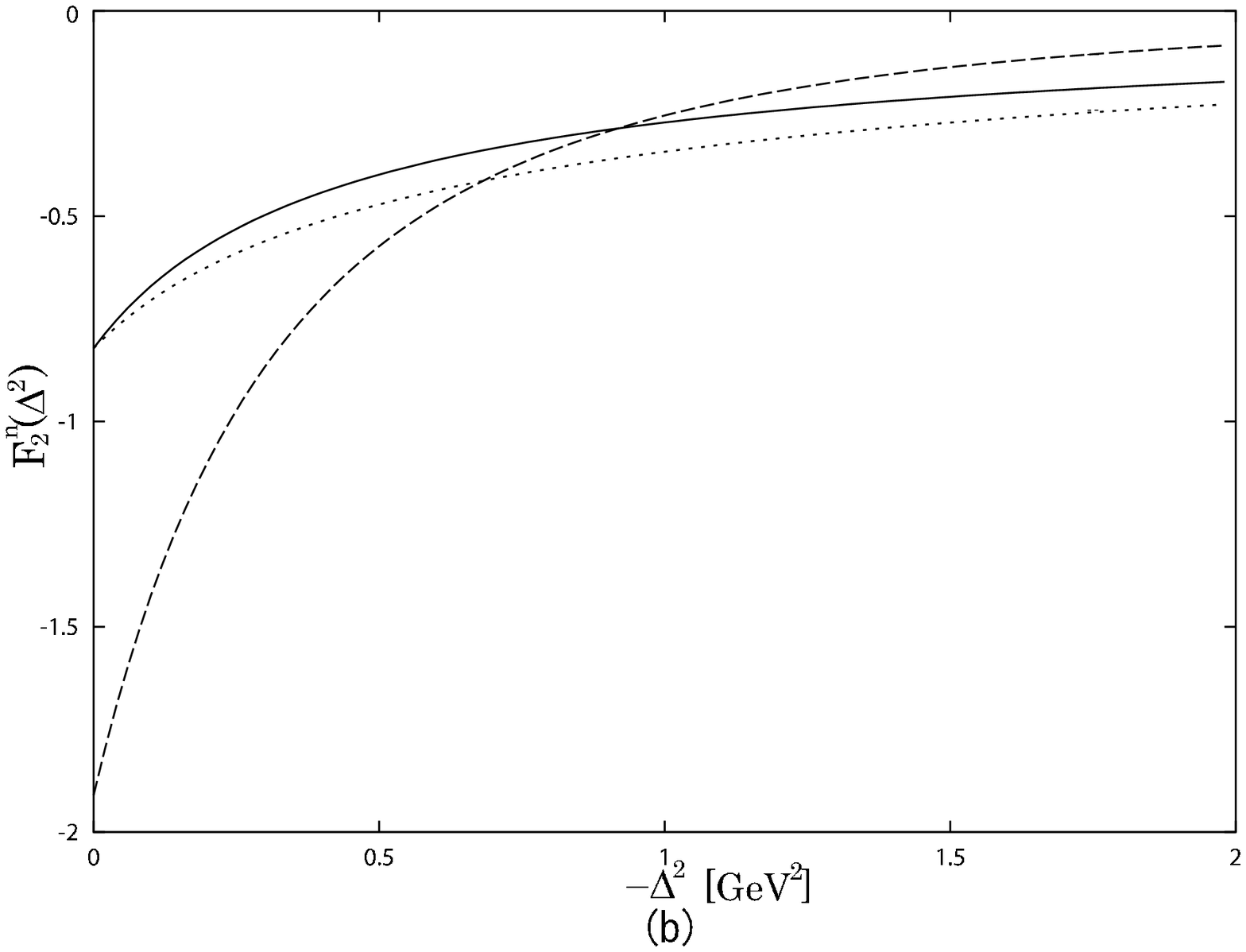,width=9cm}
\end{center}
\caption{(a) Proton form factor
$F_1^n(\Delta^2)$,
 (b)  Neutron form factor $F_2^n(\Delta^2)$.
Notation same as in Fig. 4.}
\end{figure}

\begin{figure}[hbtp]
\begin{center}
\epsfig{file=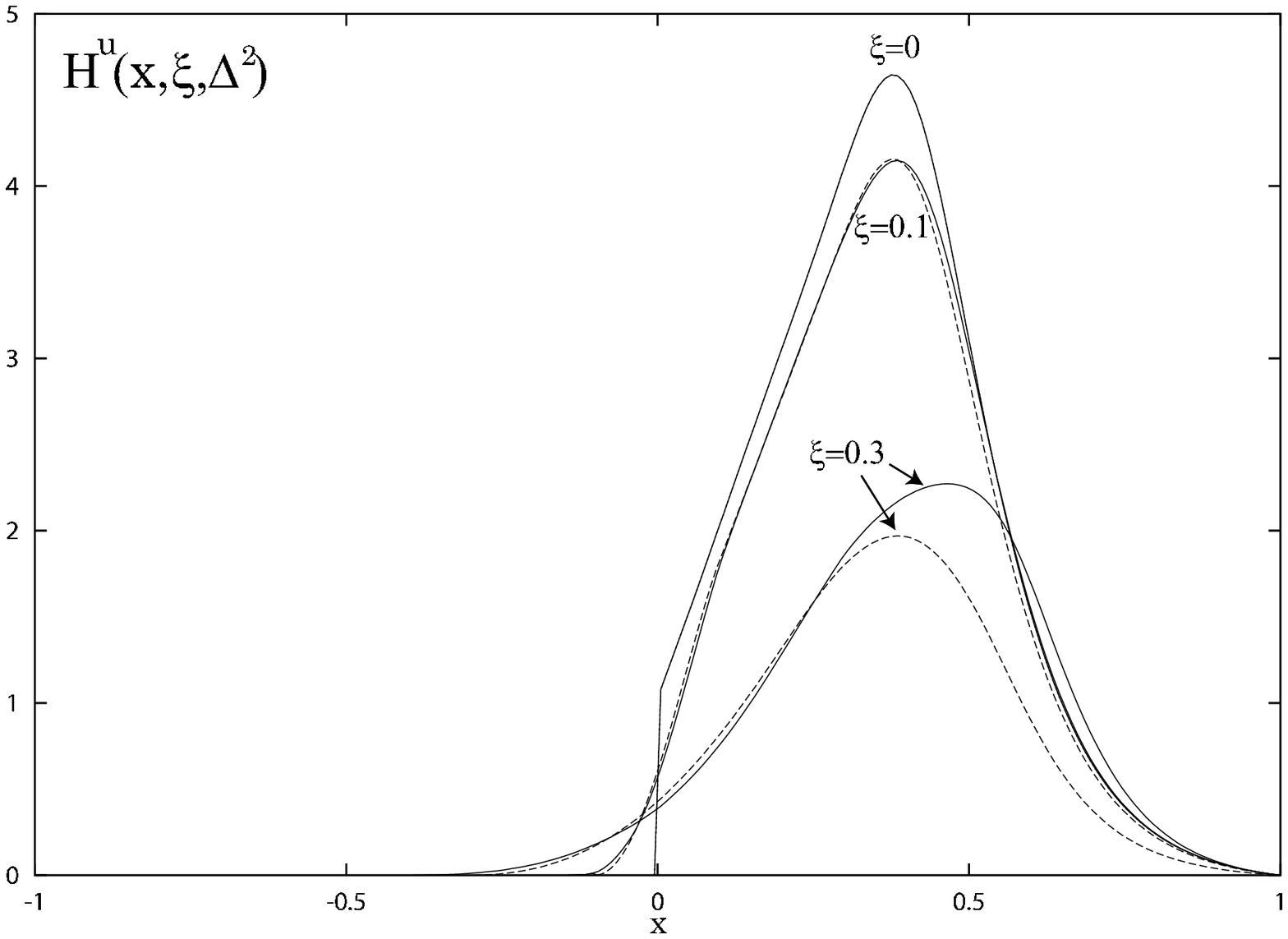,width=10cm}
\end{center}
\caption{$H^u(x,\xi,\Delta^2)$ for $\xi=0,0.1,0.3$. The
solid lines are NJL model results, and the dashed lines are
obtained using the Radyushkin's ansatz for the input
forward quark distributions calculated in NJL model.}
\end{figure}

\begin{figure}[hbtp]
\begin{center}
\epsfig{file=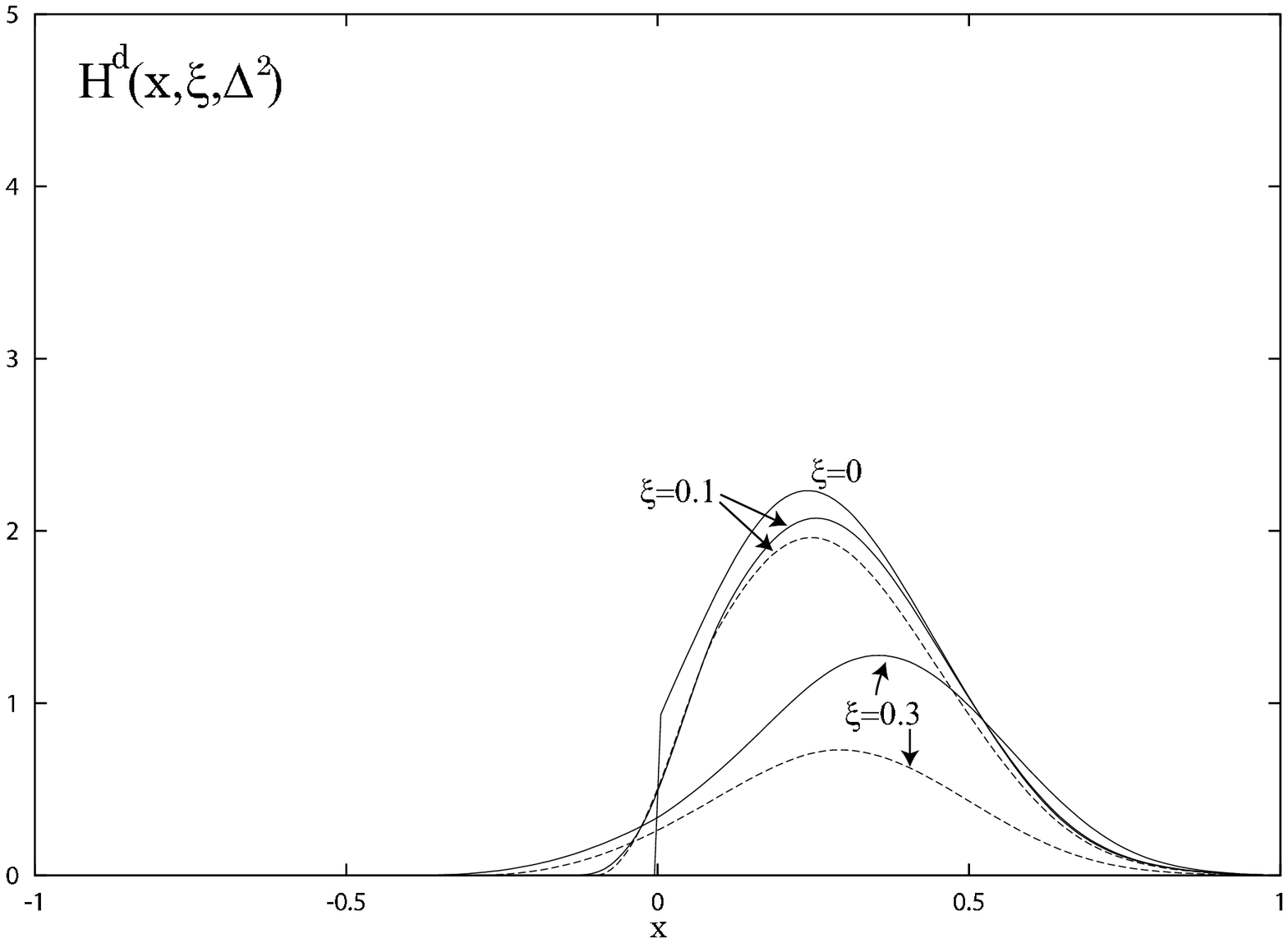,width=10cm}
\end{center}
\caption{$H^d(x,\xi,\Delta^2)$ for $\xi=0, 0.1, 0.3$.
Notation same as in Fig. 6.}
\end{figure}

\begin{figure}[hbtp]
\begin{center}
\epsfig{file=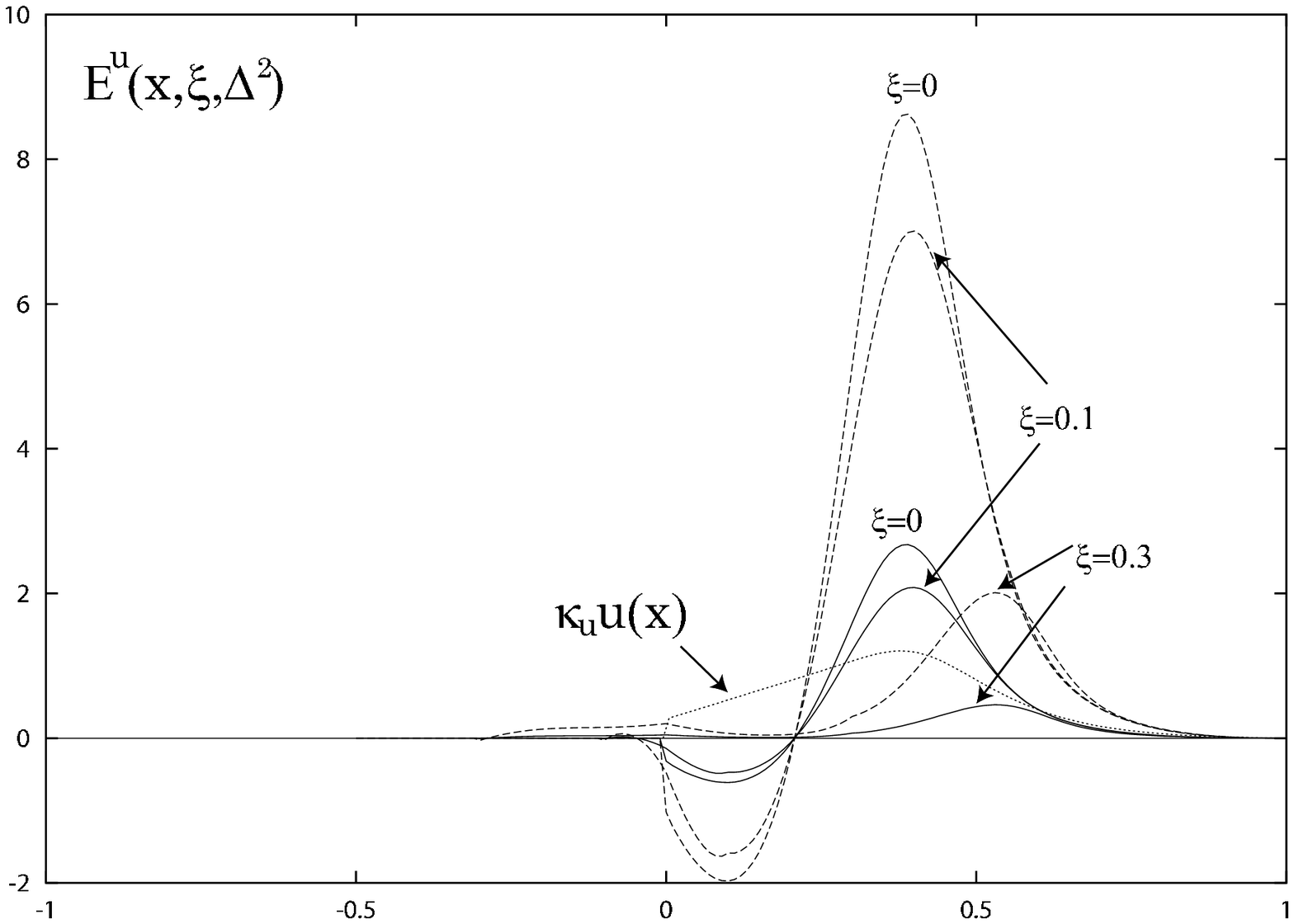,width=10cm}
\end{center}
\caption{$E^{u}(x,\xi,\Delta^2)$ for $\xi=0, 0.1, 0.3$. 
Solid lines show NJL results while
the dashed lines give the  NJL results multiplied with
a factor of 
$F_{2,EXP}^{u} (\Delta^2)/F_{2,NJL}^{u} (\Delta^2)$
(see  text for explanation).
The dotted line represent $\kappa_{u}u(x)$.}
\end{figure}

\begin{figure}[hbtp]
\begin{center}
\epsfig{file=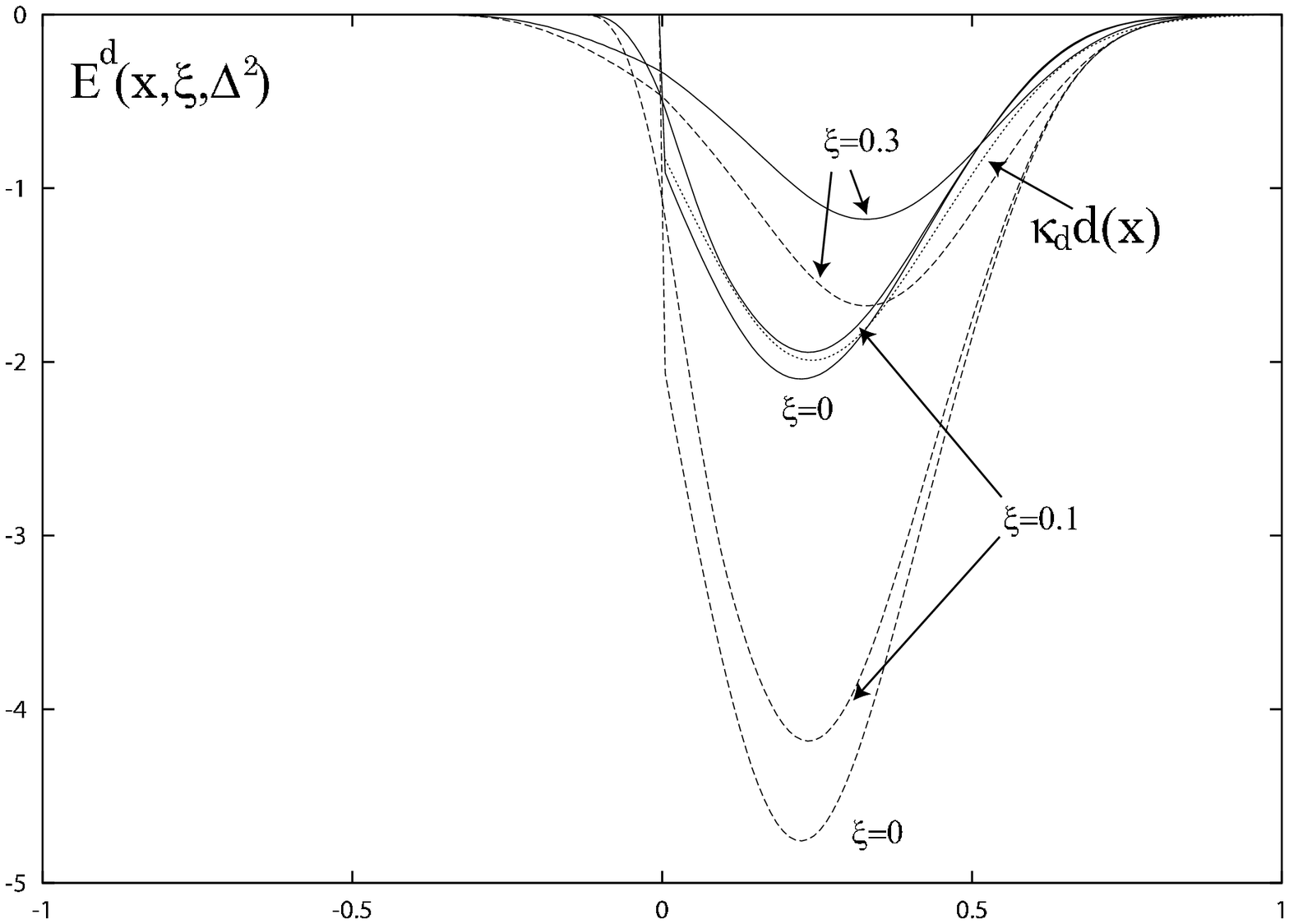,width=10cm}
\end{center}
\caption{$E^d(x,\xi,\Delta^2)$ for $\xi=0,0.1,0.3$;
Notation  same as in Fig. 8.}
\end{figure}

Having fixed the model parameters, we now present the main
results of this work. The calculated GPD's,
$H^u(x,\xi,\Delta^2), H^d(x,\xi,\Delta^2),
E^u(x,\xi,\Delta^2),$ and $E^d(x,\xi,\Delta^2)$, are
plotted in Figs. (6-9), for three different values of
$\xi=0, 0.1, 0.3$, with $\Delta^2 $ given by Eq. (\ref{Delta and xi}).  
For simplicity we have assumed
$\vec{\Delta}_{\perp}=\vec{0}_{\perp}$.

As mentioned in Section 3, the quark-current contribution
$J^Q$ to the GPDs can be decomposed into two terms, $J^{Q,P}$
and $J^{Q,C}$, corresponding respectively to the "pole-term"
and "contact term" in the diquark t-matrix, see Eq. (\ref{J^Q}). It has been
found that the contact term violates PCAC by as much as
13\% \cite{Mineo04} which is related to the use of "static
approximation" for the Faddeev vertex function. Moreover, the 
contact term contributes only in the
quark-antiquark region, $-\xi<x<\xi$, producing unphysical
kinks at $x=\pm\xi$. In view of these facts  which indicate
that the contact terms can not be assessed reliably in the
static approximation, we have chosen to
leave out the contact term contribution in our results.

Like all constituent quark models, there are no intrinsic
anti-quarks in the NJL model, therefore
$H^q(x,\xi=0,\Delta^2)$ and $E^q(x,\xi=0,\Delta^2)$
($q=u,d$) vanish  for negative x. However, unlike the
constituent quark models, the NJL model is field theoretic
in nature and the Fock states with antiquarks can appear in the
intermediate states. Accordingly, the quark-antiquark contribution
to GPDs in the region $-\xi<x<\xi$ is accessible in our
calculation.

As mentioned before, the calculated electromagnetic form
factors $F^{p,n}_2(\Delta^2)$ do not reproduce the
experimental data in the low momentum transfer region
$-\Delta^2<1$ GeV$^2$. These discrepancies would affect the
quality of $E^q(x,\xi,\Delta^2)$ calculated in our model
because $F^q_2(\Delta^2)$ is related to the first moment of
$E^q(x,\xi,\Delta^2)$ through Eq. (\ref{sum rule2}).
%
%
Consequently, we scale up our calculated
$E^q$ values by a factor of $F^{u,d(exp)}_{2}/F^{u,d(NJL)}_{2}$
and plot them in Figs. 8 and 9.

It is interesting to compare our results with those obtain
using  Radyushkin's ansatz \cite{RAD}.  Radyushkin
proposed to write the GPD in terms of a ``double
distribution" $F^q(\beta,\alpha,\Delta^2)$ which is assumed
to be factorized:
\be
F^q(\beta,\alpha,\Delta^2)=h(\alpha,\beta)q(\alpha)F_1^q
(\Delta^2)/F_1^q(0), \ee
where $q(x)$ is the forward quark distribution (or quark
distribution function) and the profile function
$h(\alpha,\beta)$ has the property of asymptotic meson
distribution amplitudes given in \cite{RAD}:
\be h(\alpha,\beta)=
\frac34\frac{(1-\beta)^2-\alpha^2}{(1-\beta)^3}.\ee
$H^q(x,\xi,\Delta^2)$ is then given by the convolution
expression:
\be H^q(x,\xi,\Delta^2)= \int_{-1}^1 d\beta
\int_{-1+|\beta|}^{1-|\beta|}d\alpha \delta(x-\beta -\alpha
\xi)F^q(\beta,\alpha,\Delta^2).\ee
Using the forward quark distributions calculated in the NJL
model as input, we plot the results obtained from the above
ansatz also in Fig. 6 and Fig. 7. We see that, in
magnitudes and in shapes, Radyushkin's ansatz gives
qualitatively similar results as the NJL model.  One
visible quantitative difference is that, as $\xi$
increases, the peak position of $H^q(x,\xi,\Delta^2)$ shifts
towards larger $x$ in the NJL model, while it stays almost
unchanged in the Radyushkin's ansatz.

In Figs. (8-9), $\kappa_q = F^q_2 (0)$. It is seen that $\kappa_d d(x)$ is 
quite similar to $E^d(x,\xi=0,\Delta^2=0)$. This is because $E^d$ receives
contribution only from the diquark current in our model. On the other hand,
our result for $\kappa_u u(x)$ is rather different from
$E^u(x,\xi=0,\Delta^2=0)$, in contrast to the results obtained with the
chiral quark-soliton model \cite{Petrov98} and constituent quark models
\cite{Boffi03,Scopetta03}.

\begin{figure}[hbtp]
\begin{center}
\epsfig{file=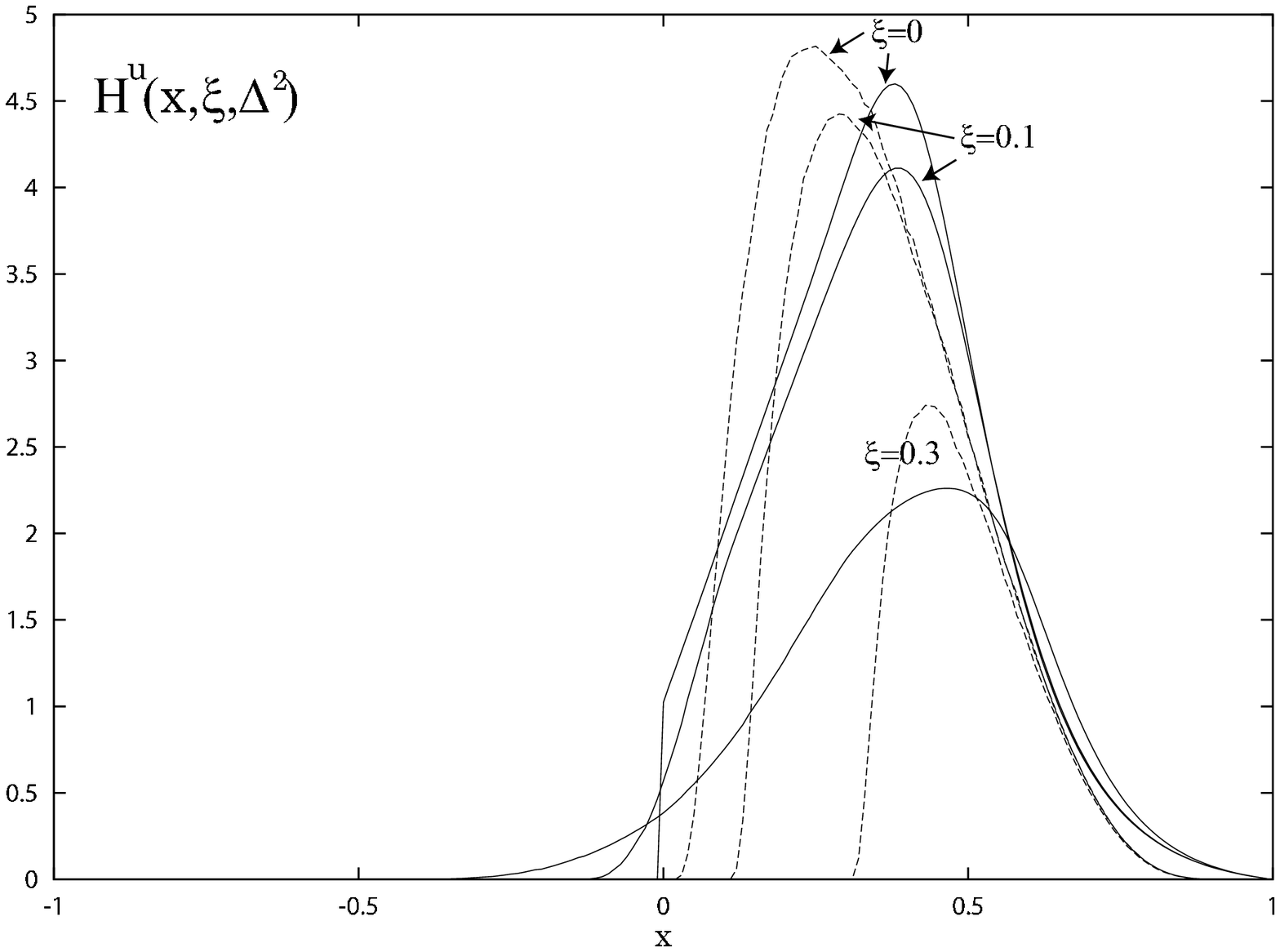,width=9cm}
\epsfig{file=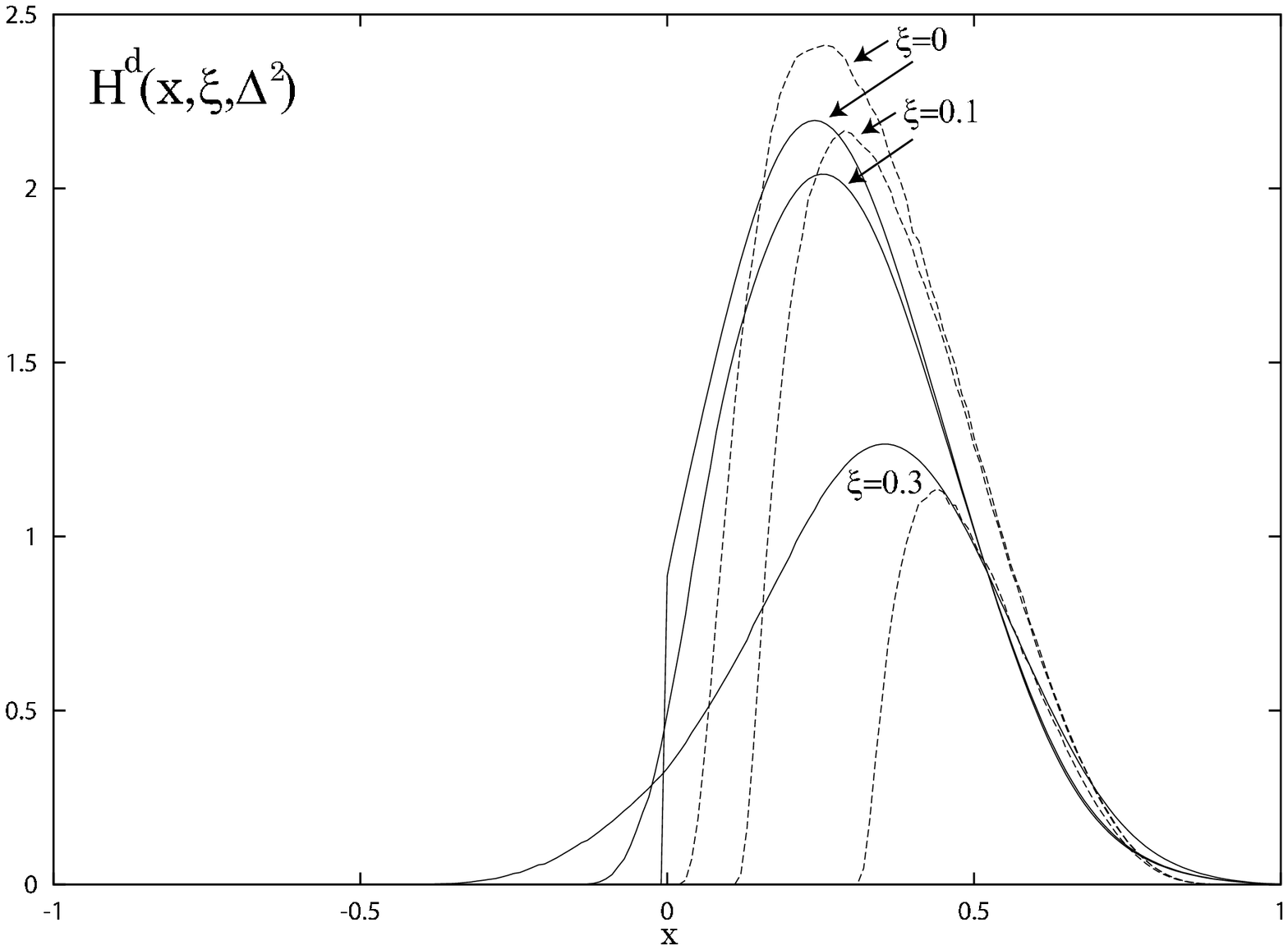,width=9cm}
\caption{(a)  $H^u(x,\xi,\Delta^2)$ for $\xi=0,0.1,0.3$.
The solid lines are the NJL model results and the dashed lines are obtained with
constituent quark models \cite{Spitzenberg04}.
(b)  $H^d(x,\xi,\Delta^2)$ for $\xi=0,0.1,0.3$. Notation same as in
   (a).}
\end{center}
\end{figure}

\begin{figure}
\begin{center}
\epsfig{file=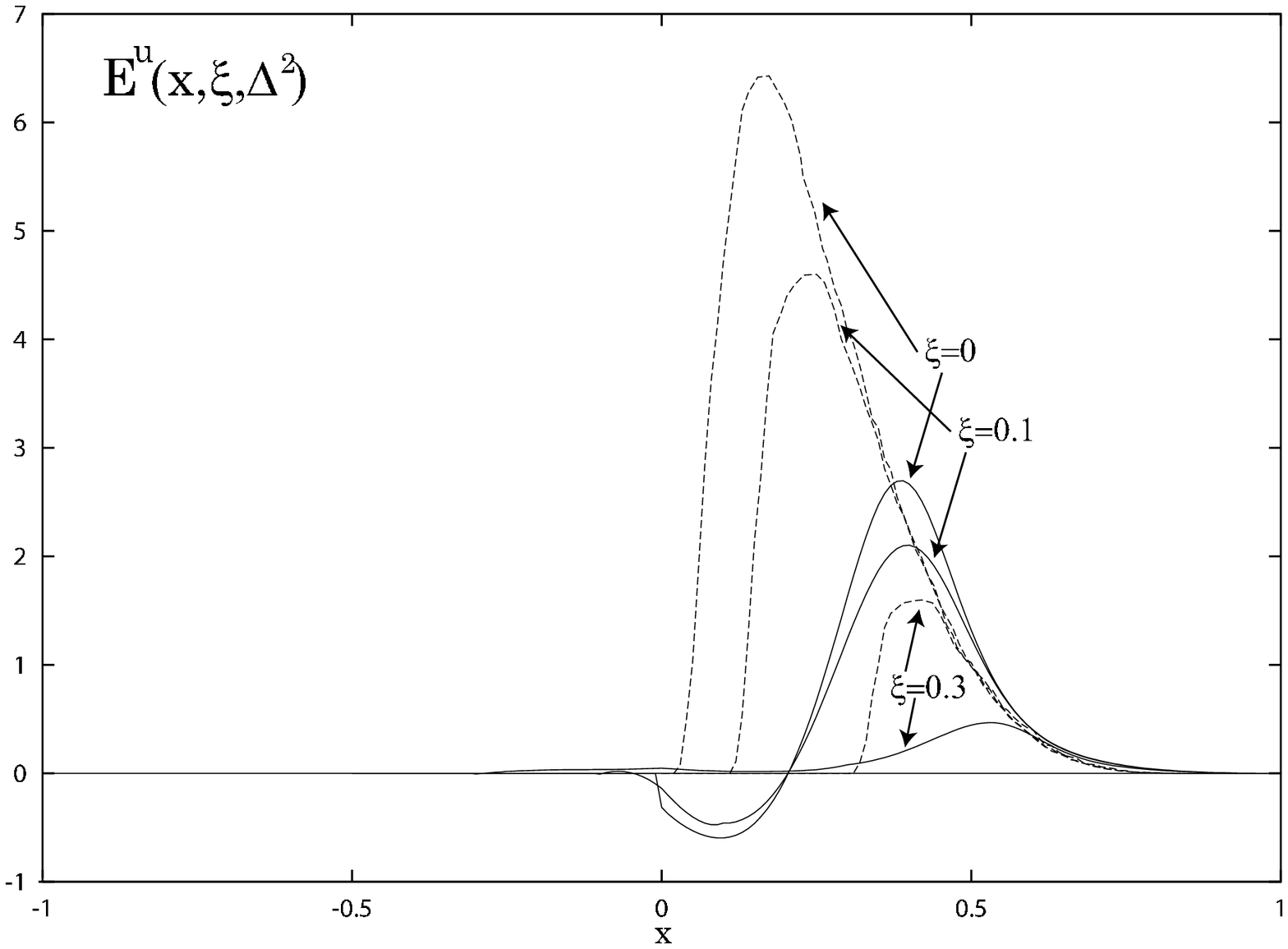,width=9cm}
\epsfig{file=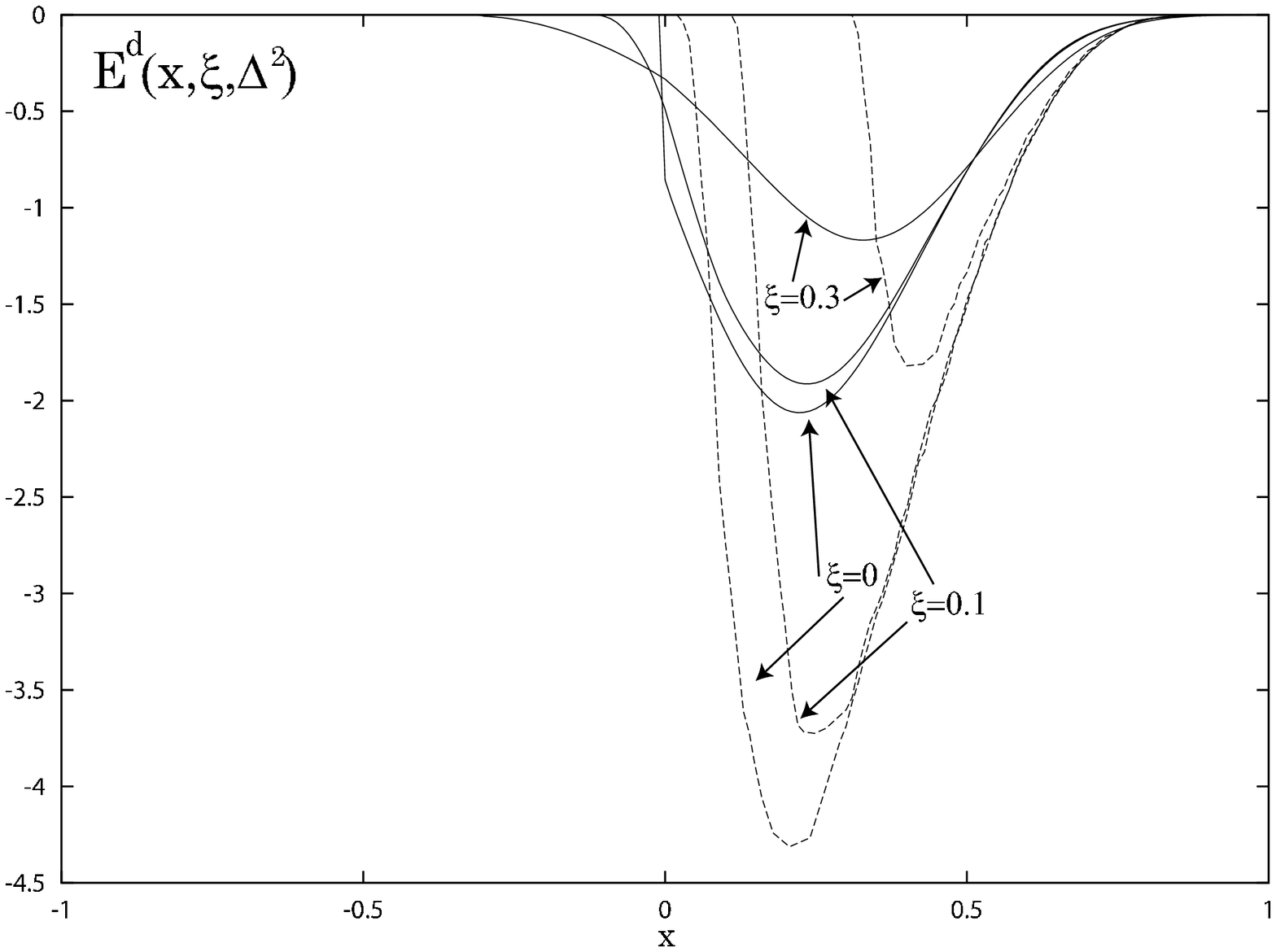,width=9cm}
\end{center}
\caption{ (a)  $E^u(x,\xi,\Delta^2)$ for $\xi=0,0.1,0.3$.
 (b)  $E^d(x,\xi,\Delta^2)$ for $\xi=0,0.1,0.3$. Notation same as in
Fig. 10.}
\end{figure}

In Figs. (10-11), we compare our results with those
obtained in a calculation using the constituent quark model
\cite{Spitzenberg04} which is calculated using a simple
gaussian wave function.
We note that their calculation of GPDs is exactly same as
\cite{Boffi03} except the use of a different wave function.
First of all, we see that the signs of the
GPDs calculated in the two models agree
except $E^u(x,\xi,\Delta^2)$,
that is, $E^u(x,\xi,\Delta^2)$ calculated in the NJL model
explicitly shows a negative contribution for small $x$.
Secondly we see
that the shifting of the peak positions towards larger $x$
with increasing $\xi$ are common in both calculations.
Finally, we observe that due to the fact that there is no
quark-antiquark contribution to the GPDs in the constituent
quark model, the curves all terminate at $x=\xi$. In
contrast, as mentioned earlier, the NJL model is field
theoretic in nature, so that quark-antiquark contributions
is non-zero in our calculation.  As a result, the range of
validity is $-\xi<x<1$ in our calculation.

Comparing our results with those obtained from the chiral
quark-soliton model \cite{Petrov98,CHR96}, we find that the
behavior in the range $-\xi<x<\xi$ is quite different. In
the case of chiral quark-soliton model, strong oscillatory behavior 
is seen around
$x=\pm\xi$, whereas our results are rather smooth. This
difference arises from the fact that in the chiral
quark-soliton model, there is a so called "d-term" contribution
\cite{Goeke01,Polyakov99,VDH03} which corresponds to the case
where the active quark and antiquark are correlated in
the scalar isoscalar (or $\sigma$) channel
\cite{Theussl04}. Such a contribution is supported by the recent
preliminary HERMES data \cite{Hermes02} on beam-charge asymmetry 
in DVCS but is not included in our model.

\section{Summary}

In this work, we have calculated the spin-averaged ($H^q$) and
helicity-flip ($E^q$) GPDs of the proton, using
the NJL model based on a relativistic Faddeev approach with
"static approximation".  The NJL model is a field theoretic
approach which has been successfully used in the studies of
the static properties and parton
distribution functions of the nucleon. 
Hence the NJL model provides a reasonable framework in
which to calculate the GPDs or off-forward parton distribution
functions. Among other things, there are two major
advantages of adopting this model. First, due to the fact that
NJL model is a relativistic field theoretic model Fock states
with anti-quarks can exist in the intermediate states,
hence quark-antiquark contributions to the GPDs in the
region $-\xi < x < \xi$ are non-vanishing.  In addition, the model
independent sum rules relating the GPDs and nucleon
electromagnetic form factors are satisfied.

The calculated GPDs are qualitatively similar to those
calculated with the Radyushkin's double distribution ansatz
with forward parton distribution functions calculated in
the NJL model as inputs.  Comparing our results with those
obtained in constituent quark models
\cite{Boffi03,Scopetta03}, we find that the general
features are similar, except for the fact that the region
$-\xi<x<\xi$ is not accessible in the latter.

In our present treatment of the NJL model, as well as in other quark models,
configurations with intrinsic antiquarks are not present.
Hence it is not possible to investigate GPDs in the region
$x<-\xi$.  In our case, antiquark contribution can be
studied if we include the pion cloud surrounding the
three-quark core. In addition, since NJL model is an effective quark theory in the low energy regime,
we need to evolve our results, according to perturbative QCD,
in order to compare them with data taken in high-energy experiments.
Such an NLO $Q^2$-evolution of the calculated GPDs, from the low-momentum
scale to the experimental one, has been carried in Refs. 
\cite{Scopetta03,Pasquini04}. 
We will leave these improvements to future investigations.

\section*{Acknowledgments}
The authors wish to thank T. Spitzenberg and M. Vanderhaeghen for helpful
discussions. This work is supported, in part, by the National Science
Council of ROC under grant Nos. NSC93-2112-M002-004, NSC93-2112-M002-058,
and the Grant in Aid for Scientific Research of the Japanese Ministry of
Education, Sports, Science and Technology, Project No. 16540267.

\newpage
\appendix{\LARGE Appendices}
\section{Matrix elements of Dirac spinors}
Table 1 contains the matrix elements of Dirac spinors used
in our calculations \cite{BRO}. The convention of
\cite{BRO} is adopted here:
\ba u_N(p,\lambda)&=&\frac{1}{\sqrt{\sqrt{2}p^+}}
(\sqrt{2}p^+ +\vec{\alpha}_{\perp}\cdot \vec{p}_{\perp}
+\beta M_N)\chi(\lambda),\\
\chi(+1)&=&\frac{1}{\sqrt{2}} \left(
  \begin{array}{c}
   1\\0\\1\\0\\
\end{array}\right), \,\,\,
\chi(-1)=\frac{1}{\sqrt{2}} \left(
  \begin{array}{c}
   0\\1\\0\\-1\\
\end{array} \right);
\ea
and $a_{\perp}(\lambda)$ and $a_{\perp} \land b_{\perp}$
are defined by
\be a_{\perp}(\lambda) \equiv \lambda a^1 + ia^2, \,\,\,
a_{\perp} \land b_{\perp} \equiv a^1 b^2 -a^2 b^1. \ee
\begin{table}[htbp]
\begin{center}
\caption{Matrix elements of Dirac spinors ${\bar u}_N
(p',\lambda'){\cal M}u_N(p,\lambda)$}
\begin{tabular}{|c|c|c|}
\hline ${\cal M}$ & $\frac{\delta_{\lambda',\lambda}}{\sqrt{p'^+ p^+}}{\bar u}_N
(p',\lambda') {\cal
M}u_N(p,\lambda)$ &
$\frac{\delta_{\lambda',-\lambda}}{\sqrt{p^+ p'^+}}{\bar u}_N(p',\lambda')
{\cal M}u_N(p,\lambda)$\\
\hline\hline 1 & $\frac{M_N}{p'^+}+\frac{M_N}{p^+}$ &
$\frac{p'_{\perp}(\lambda)}{p'^+}-
\frac{p_{\perp}(\lambda)}{p^+}$ \\
$\gamma^+$ & 2 & 0 \\
$\gamma^-$ & $\frac{1}{p'^+ p^+} (\vec{p'}_{\perp}\cdot
\vec{p}_{\perp}+M_N^2 +i\lambda p'_{\perp}\land p_{\perp})$
& $\frac{M_N}{p'^+ p^+}( p'_{\perp}(\lambda)-p_{\perp}(\lambda))$\\
$\vec{\gamma}_{\perp}\cdot \vec{a}_{\perp}$ &
$\vec{a}_{\perp}\cdot \left(\frac{p'_{\perp}}{p'^+}+
\frac{p_{\perp}}{p^+}\right) - i\lambda a_{\perp}\land
\left(\frac{p'_{\perp}}{p'^+}-
\frac{p_{\perp}}{p^+}\right)$ &
$-a_{\perp}(\lambda)\left(\frac{M_N}{p'^+}-
\frac{M_N}{p^+}\right)$\\
$\gamma^- \gamma^+$ & $\frac{2}{p'^+}M_N$&
$\frac{2}{p'^+}p'_{\perp}(\lambda)$\\
$\vec{\gamma}_{\perp}\cdot \vec{a}_{\perp}\gamma^+ $
& 0 & $2a_{\perp}(\lambda)$\\
$\gamma^- \gamma^+ \gamma^-$&$\frac{2}{p'^+p^+}
(\vec{p'}_{\perp}\cdot \vec{p}_{\perp}+M_N^2+i\lambda
p'_{\perp}\land p_{\perp})$
&$\frac{2}{p'^+p^+}(p'_{\perp}(\lambda)-p_{\perp}(\lambda))$\\
$\gamma^- \gamma^+ \vec{\gamma}_{\perp}\cdot
\vec{a}_{\perp}$&$\frac{2}{p'^+}(\vec{a}_{\perp}\cdot
\vec{p'}_{\perp} -i\lambda a_{\perp}\land p'_{\perp})$
&$-\frac{2M_N}{p'^+}a_{\perp}(\lambda)$\\
$\gamma^+ \gamma^- \vec{\gamma}_{\perp}\cdot
\vec{a}_{\perp}$&$\frac{2}{p^+}(\vec{a}_{\perp}\cdot
\vec{p}_{\perp} +i\lambda a_{\perp}\land p_{\perp})$
&$\frac{2M_N}{p^+}a_{\perp}(\lambda)$\\
$\vec{a}_{\perp}\cdot \vec{\gamma}_{\perp}\gamma^+
\vec{b}_{\perp}\cdot \vec{\gamma}_{\perp}$&
$2(\vec{a}_{\perp}\cdot \vec{b}_{\perp}+i\lambda a_{\perp}\land b_{\perp})$&0\\
\hline

\end{tabular}
\end{center}
\end{table}

\newpage
\section{Quark current contribution}

In Eq. (\ref{QC}), the $K^+$-integral can be trivially
performed.  Then with the help of table 1, we get
\ba &&{\bar u}_N(p',\lambda')(\fslash{k'}+M)\gamma^+
(\fslash{k}+M)u_N(p,\lambda)  \nonumber\\
&=& {\bar u}_N(p',\lambda)u_N(p,\lambda) \frac{P^+}{M_N}
\left[ (x^2-\xi^2)M_N^2 +(2x+\xi -\xi^2 )MM_N
-\frac{(1-x)^2}{4}
\vec{\Delta}_{\perp}^2 \right. \nonumber\\
&&+(1-\xi^2)(\vec{K}_{\perp}^2 +M^2) \left.
+(1+x+\xi+\xi^2)\vec{K}_{\perp}\cdot \vec{\Delta}_{\perp}+
i\lambda(1+2\xi +x){K}_{\perp}\land{\Delta}_{\perp}
\right]\nonumber\\
&&(\mbox{for $\lambda'=\lambda$})\nonumber\\
&=&{\bar u}_N(p',-\lambda)u_N(p,\lambda) \frac{P^+}{M_N}
\left[2(1-x)M_N \left( (x+2\xi
\frac{K_{\perp}(\lambda)}{\Delta_{\perp}(\lambda)})M_N +M
\right) \right]\nonumber\\
&&(\mbox{for $\lambda'=-\lambda$}) \nonumber \ea

The rest of Eq. (\ref{QC})is given by
\be -Z_N \int \frac{d^2 K_{\perp} }{(2\pi)^4}dK^-
\frac{\tau_D^C(p-k)+\tau_D^P(p-k)}{(k'^2-M^2)(k^2-M^2)}. \ee
After performing the $K^-$-integral, we obtain
%
%
\ba
&& \int dK^- \frac{-\tau_D^C (p-k)}{(k'^2-M^2)(k^2-M^2)} \equiv
 F_C(x,\xi,\vec{K}_{\perp},\vec{\Delta}_{\perp},M^2,M_D^2)\nonumber\\
&=& \frac{-4G_s\theta(-\xi<x<\xi)}
{B_Q(x,\xi,\vec{K}_{\perp},\vec{\Delta}_{\perp},M^2,M_D^2)}
,\nonumber\\
&& \int dK^- \frac{-\tau_D^P (p-k)}{(k'^2-M^2)(k^2-M^2)} \equiv
 F_P(x,\xi,\vec{K}_{\perp},\vec{\Delta}_{\perp},M^2,M_D^2)\nonumber\\
&=& \frac{g_D^2}{2A_Q(x,-\xi,\vec{K}_{\perp},-\vec{\Delta}_{\perp},M^2,M_D^2)}\nonumber\\
&\times&
\left( \frac{(1-x)\theta (x>\xi)}{A_Q(x,\xi,\vec{K}_{\perp},\vec{\Delta}_{\perp},M^2,M_D^2)}
+
\frac{(x+\xi)\theta(-\xi<x<\xi)}{B_Q(x,\xi,\vec{K}_{\perp},\vec{\Delta}_{\perp},M^2,M_D^2)}
\right),\nonumber\\
\ea

where
\ba
A_Q(x,\xi,\vec{K}_{\perp},\vec{\Delta}_{\perp},M^2,M_D^2)
&=&
(1-x)(x-\xi)(1+\xi)P^2-(x-\xi)(\vec{K}_{\perp}^2+M_D^2)\nonumber\\
&&-(1-x) [
(\vec{K}_{\perp}+\frac{\vec{\Delta}_{\perp}}{2})^2+M^2],
\nonumber \ea
\be B_Q(x,\xi,\vec{K}_{\perp},\vec{\Delta}_{\perp},M^2) =
2\left[ \xi(x^2-\xi^2)P^2 -\xi (\vec{K}_{\perp}^2+M^2+
\frac{\Delta_{\perp}^2}{4})-x \vec{K}_{\perp}\cdot
\vec{\Delta}_{\perp} \right].\nonumber \ee

Combining the above results and with the help of
Eq. (\ref{helicity relation}), we arrive at the final
expressions for $H^Q$ and $E^Q$ which can be decomposed into
the 'pole' and 'contact' term contributions (Eq. (\ref{J^Q})):
$H^Q=H^C+H^P,E^Q=E^C+E^P$.
In the following we will
explicitly write down the results of $H^{C,P}$ and $E^{C,P}$ with
PV regularization scheme (Eq. (\ref{PV})):
\ba &&H^{C,P}(x,\xi,\Delta^2)\nonumber\\
&=& \sum_i c_i \int \frac{d^2 K_{\perp}}{(2\pi)^3} \Biggr\{
\left[x^2-\xi^2+2\xi^2(1-x) \left(x+2\xi
\frac{K_{\perp}(\lambda)}{\Delta_{\perp}(\lambda)}
\right)
\right]M_N^2+2x(1-\xi^2 )MM_N \nonumber\\
&&- \frac{ (1-x)^2}{4}\vec{\Delta}_{\perp}^2 +
(1-\xi^2)(\vec{K}_{\perp}^2 +M_i^2)
+(1+x+\xi+\xi^2)\vec{K}_{\perp}\cdot
\vec{\Delta}_{\perp}\Biggr\}\nonumber\\
&\times&\frac{F_{C,P}(x,\xi,\vec{K}_{\perp},\vec{\Delta}_{\perp},
M_i^2,{M_D}_i^2)}{(1-\xi^2)}, \label{H_Q}\\
&&E^{C,P}(x,\xi,\Delta^2)\nonumber\\
&=& \sum_i c_i \int \frac{d^2 K_{\perp}}{(2\pi)^3}
2(1-x)M_N\left[ \left(x+2\xi
\frac{K_{\perp}(\lambda)}{\Delta_{\perp}(\lambda)}
\right)M_N+M\right]
F_{C,P}(x,\xi,\vec{K}_{\perp},\vec{\Delta}_{\perp},M_i^2,{M_D}_i^2),
\nonumber\\
\label{E_Q} \ea
where $M_i^2=M^2+\Lambda_i^2$, $M_{Di}^2=M_D^2+\Lambda_i^2$
and the $c_i$'s are given in Eq. (\ref{PV}) and
Eq. (\ref{PVform}).

\newpage
\section{Diquark current contribution}

The diquark current contribution is given in
Eq. (\ref{F_D}).  In order to simply the calculation, we
shall assume the initial and final diquarks are on shell,
that is, $t'^2=t^2=M_D^2$. Then $T\cdot \Delta_D =0$, and
$T^2=M_D^2-\Delta_D^2/4$. In the above and later
discussions we will explicitly distinguish $\Delta^{\mu}$
and $\Delta_D^{\mu}$, since under the on-shell diquark
approximation the frame where we calculate diquark GPDs is
not necessarily the same as the one originally chosen for
the nucleon GPDs, so that in general
$\Delta^{\mu}\ne\Delta_D^{\mu}$. Thus, Eq. (\ref{F_D})
becomes
\ba &&\left(F^D_s (y,\zeta,\Delta_D^2) T^+ + F^D_a
(y,\zeta,\Delta_D^2)\Delta^+ \right)\nonumber\\
&=& ig_D^2 \int\frac{d^4 K}{(2\pi)^4} \delta \left(
y-\frac{K^+}{T^+} \right) tr\left[S(k')\gamma^+ S(k)S(T-K)
\right], \label{F_Dsa} \ea
where $\Delta_D^+ =-2\zeta T^+$, and in the frame where
$\vec{T}_{\perp}=\vec{0}_{\perp}$, $\Delta_D^2$ is given by
$\Delta_D^2= -\frac{4\zeta^2 M_D^2 + \vec{\Delta}_{D
\perp}^2}{1-\zeta^2}$.

Integrating Eq. (\ref{F_Dsa}) over $y$, we can reproduce the
diquark form factors $G_{s,a}^D$,
\be \int_{-1}^{1} dy
F^D_{s,a}(y,\zeta,\Delta_D^2)=G_{s,a}^D (\Delta_D^2), \ee
where we see that $G_{s,a}^D(\Delta_D^2)$ is independent of
$\zeta$ as required. Furthermore for on-shell diquarks, due
to the symmery under the exchange of $t$ and $t'$, we
explicitly find that $G_a^D(\Delta_D^2)=0$.

After integrating over $K^+$ and $K^-$, we obtain
\ba &&\left(F^D_s (y,\zeta,\Delta_D^2)-2\zeta F^D_a
(y,\zeta,\Delta_D^2) \right)T^+ \nonumber\\
&=& 6g_D^2 \sum_i c_i \int\frac{d^2 K_{\perp}}{(2\pi)^3}
\Bigg[  \nonumber\\
& &\theta(y>\zeta)  \frac{ \zeta^2(1-y)^2 T^2
+(1-\zeta^2)(\vec{K}_{\perp}^2+M_i^2)+ \frac{(1-y)^2
\Delta_D^2}{4} +\zeta(1-y) \vec{K}_{\perp}\cdot
\vec{\Delta}_{D\perp}}
{A_D(y,\zeta,\vec{K}_{\perp},\vec{\Delta}_{D\perp},M_i^2)
A_D(y,-\zeta,\vec{K}_{\perp},-\vec{\Delta}_{D\perp},M_i^2)}
\nonumber\\
&-&\theta(|y|<\zeta)\nonumber\\
&\times&\frac{ (\zeta^2(\zeta-y)
+y^2(1-\zeta))T^2+(1+\zeta) (\vec{K}_{\perp}^2+M_i^2)+
\frac{(1-y)\Delta_D^2+(\zeta-y)\vec{\Delta}_{D\perp}^2}{4}
+y \vec{K}_{\perp}\cdot\vec{\Delta}_{D\perp}}
{A_D(y,\zeta,\vec{K}_{\perp},\vec{\Delta}_{D\perp},M_i^2)
B_D(y,\zeta,\vec{K}_{\perp},\vec{\Delta}_{D\perp},M_i^2)},
\nonumber\\
\ea
with
\ba &&
A_D(y,\zeta,\vec{K}_{\perp},\vec{\Delta}_{D\perp},M_i^2)
\nonumber\\
&=& (y+\zeta)(y-1)(1-\zeta)T^2 +(y+\zeta)\vec{K}_{\perp}^2
-(y-1)(\vec{K}_{\perp}-\frac{\vec{\Delta}_{D\perp}}{2})^2
+(1+\zeta)M_i^2,\nonumber\\
&& B_D(y,\zeta,\vec{K}_{\perp},\vec{\Delta}_{D\perp},M_i^2)
\nonumber\\
&=& 2\zeta (y-\zeta)T^2
 +\frac{y-\zeta}{y+\zeta}
(\vec{K}_{\perp}-\frac{\vec{\Delta}_{D\perp}}{2})^2
-(\vec{K}_{\perp}+\frac{\vec{\Delta}_{D\perp}}{2})^2
-\frac{2\zeta}{y+\zeta} M_i^2,\nonumber \ea where
$T^2=M_D^2-\Delta_D^2/4$.

Next we need to calculate the following integral
\ba && F_{\lambda',\lambda}^{D/N}
(z,\xi,\vec{T}_{\perp},\vec{
\Delta}_{\perp})\nonumber\\
 &=& \frac{1}{2\pi}\int {dT^+ dT^-}
{\cal F}_{\lambda',\lambda}^{D/N}(z,\xi,T,\Delta)
\nonumber\\
&=& \frac{ig_D^2 Z_N}{2\pi G_s^{D}(\Delta^2)}{\bar
u}(p',\lambda') \int dT^+ dT^- \frac{\delta
(z-T^+/P^+)S(P-T)}
{(t'^2-M_D^2)(t^2-M_D^2)}u(p,\lambda).\nonumber\\
\ea
Insert the PV-regularization scheme, and following the same
steps as indicated in appendix B, we get
\ba
&&F_{\lambda',\lambda}^{D/N}(z,\xi,\vec{T}_{\perp},\vec{
\Delta}_{\perp}) = \frac{g_D^2 Z_N}{2G_s^D(\Delta^2)} {\bar
u}(p',\lambda') u(p,\lambda)
\sum_i c_i \Biggr[\nonumber\\
&&
\frac{\theta(z>\xi)(1-z)(M+M_N-B_{D/N}^{\lambda',\lambda}
(T^-=P^--\frac{\vec{T}_{\perp}^2+M_i^2}{2(1-z)P^+},z,\vec{T}_{\perp},
\vec{\Delta}_{\perp}))}
{A_{D/N}(z,\xi,\vec{T}_{\perp},\vec{\Delta}_{\perp},M_i^2,M_{Di}^2)
A_{D/N}(z,-\xi,\vec{T}_{\perp},-\vec{\Delta}_{\perp},M_i^2,M_{Di}^2)}
\nonumber\\
&-& \frac{\theta(|z|<\xi)(x+\xi)
(M+M_N-B_{D/N}^{\lambda',\lambda} (T^-=\xi
P^-+\frac{(\vec{T}_{\perp}-\vec{\Delta}_{\perp}/2)^2+M_{Di}^2}{2(z+\xi)P^+}
,z,\vec{T}_{\perp},\vec{\Delta}_{\perp}))}
{A_{D/N}(z,\xi,\vec{T}_{\perp},\vec{\Delta}_{\perp},M_i^2,M_{Di}^2)
C_{D/N}(z,\xi,\vec{T}_{\perp},\vec{\Delta}_{\perp},M_{Di}^2)}
\Biggl],\nonumber\\
\ea
with
\ba && A_{D/N}(z,\xi,\vec{T}_{\perp},\vec{\Delta}_{\perp},
M_i^2,M_{Di}^2) \nonumber\\
&=& (1-z)(z+\xi)(1-\xi)P^2
-(z+\xi)(\vec{T}_{\perp}^2+M_i^2)
-(1-z)((\vec{T}_{\perp}-\vec{\Delta}_{\perp}/2)^2+M_{Di}^2),
\nonumber\\
&&
C_{D/N}(z,\xi,\vec{T}_{\perp},\vec{\Delta}_{\perp},M_{Di}^2)
\nonumber\\
&=& 2\xi( (z^2-\xi^2)P^2
-\vec{T}_{\perp}^2-\vec{\Delta}_{\perp}^2/4
-M_{Di}^2)-2z\vec{T}_{\perp}\cdot \vec{\Delta}_{\perp},
\nonumber\\
&& B_{D/N}^{\lambda',\lambda}
(T^-,z,\vec{T}_{\perp},\vec{\Delta}_{\perp}))\nonumber\\
&=& \frac{\delta_{\lambda',\lambda}}{2M_N}
[z(M_N^2-\vec{\Delta}_{\perp}^2/4)+2(1-\xi^2)T^-P^+ +\xi
\vec{T}_{\perp}\cdot \vec{\Delta}_{\perp}-i\lambda
T_{\perp}\land\Delta_{\perp}]
\nonumber\\
&+& \delta_{\lambda',-\lambda}M_N
\left(z+2\xi\frac{T_{\perp}(\lambda)}
{\Delta_{\perp}(\lambda)}\right),\nonumber \ea
where $P^2=M_N^2-\Delta^2/4$.

With help of Eq. (\ref{helicity relation}), we finally
arrive at
\ba H^D(x,\xi,\Delta^2) &=& \int dy \int dz \delta(x-yz)
\frac{F_s^D(y,\zeta,\Delta_D^2)-2\zeta
F_a^D(y,\zeta,\Delta_D^2)}{G_s^D(\Delta^2)}\nonumber\\
&&\times\int \frac{d^2 T_{\perp}}{(2\pi)^3}
H^{D/N}(z,\xi,\vec{T}_{\perp},\vec{\Delta}_{\perp}),
\nonumber\\
E^D(x,\xi,\Delta^2) &=&\int dy \int dz \delta(x-yz)
\frac{F_s^D(y,\zeta,\Delta_D^2)-2\zeta
F_a^D(y,\zeta,\Delta_D^2)}{G_s^D(\Delta^2)}\nonumber\\
&&\times\int \frac{d^2 T_{\perp}}{(2\pi)^3}
E^{D/N}(z,\xi,\vec{T}_{\perp},\vec{\Delta}_{\perp}),
\nonumber\\
\label{H_D} \ea
where
\ba &&H^{D/N}(z,\xi,\vec{T}_{\perp},\vec{\Delta}_{\perp})
\nonumber\\
&=& \frac{g_D^2 Z_N}{2}\sum_i c_i \Biggr[
\frac{\theta(z>\xi)(1-z)}{
A_{D/N}(z,\xi,\vec{T}_{\perp},\vec{\Delta}_{\perp})
A_{D/N}(z,-\xi,\vec{T}_{\perp},-\vec{\Delta}_{\perp})}
\nonumber\\
&+& \frac{\theta(|z|<\xi)(z+\xi)}{
A_{D/N}(z,\xi,\vec{T}_{\perp},\vec{\Delta}_{\perp})
C_{D/N}(z,\xi,\vec{T}_{\perp},\vec{\Delta}_{\perp})}
\Biggl]\tilde{H}^{D/N}(T_i^-,z,\xi,\vec{T}_{\perp},\vec{\Delta}_{\perp}),\nonumber\\
\ea
\ba &&E^{D/N}(z,\xi,\vec{T}_{\perp},\vec{\Delta}_{\perp})\nonumber\\
&=& \frac{Z_N}{2}\sum_i c_i
\Biggr[\frac{\theta(z>\xi)(1-z)}{
A_{D/N}(z,\xi,\vec{T}_{\perp},\vec{\Delta}_{\perp})
A_{D/N}(z,-\xi,\vec{T}_{\perp},-\vec{\Delta}_{\perp})}
\nonumber\\
&-& \frac{\theta(|z|<\xi)(z+\xi)}{
A_{D/N}(z,\xi,\vec{T}_{\perp},\vec{\Delta}_{\perp})
C_{D/N}(z,\xi,\vec{T}_{\perp},\vec{\Delta}_{\perp})}
\Biggl]\tilde{E}^{D/N}(T_i^-,z,\xi,\vec{T}_{\perp},\vec{\Delta}_{\perp}),\nonumber\\
\ea
in which
\ba
&& \tilde{H}^{D/N}(T_i^- ,z,\xi,\vec{T}_{\perp},\vec{\Delta}_{\perp})\nonumber\\
&=& \frac{zM_N}{1-\xi^2} \Biggr[ (1+\xi^2)(M+M_N)-\xi^2 M_N
(z+2\xi
\frac{K_{\perp}(\lambda)}{\Delta_{\perp}(\lambda)}
)\nonumber\\
&-& \frac{z(M_N^2-\vec{\Delta}_{\perp}^2/4)+2(1-\xi^2)T_i^-
P^+ +\xi\vec{T}_{\perp}\cdot \vec{\Delta}_{\perp}}{2M_N}
\Biggl],\nonumber\\
\ea
with
\be T_i^-=\theta(z>\xi)(P^-
-\frac{\vec{T}_{\perp}^2+M_i^2}{2(1-z)P^+})
+\theta(|z|<\xi)(\xi P^- + \frac{(\vec{T}_{\perp}
-\vec{\Delta}_{\perp}/2)^2+M_{Di}^2}{2(z+\xi)P^+}), \ee
and
\be \tilde{E}^{D/N}(T_i^-
,z,\xi,\vec{T}_{\perp},\vec{\Delta}_{\perp})= zM_N\left(
M+M_N-M_N(z+2\xi
\frac{K_{\perp}(\lambda)}{\Delta_{\perp}(\lambda)}
)\right). \ee

Note that if we integrate Eq. (\ref{H_D}) over $x$ with
$\Delta^2=\Delta_D^2$ and fixed $\zeta$, then we can
reproduce the diquark current contributions to the form
factors,
\ba \int_{-1}^1 dx H^D(x,\xi,\Delta^2) &=& \int_{-\xi}^1 dz
\int \frac{d^2 T_{\perp}}{(2\pi)^3}
H^{D/N}(z,\xi,\vec{T}_{\perp},\vec{\Delta}_{\perp})=
F_1^D(\Delta^2),\nonumber\\
\int_{-1}^1 dx E^D(x,\xi,\Delta^2) &=& \int_{-\xi}^1 dz
\int \frac{d^2 T_{\perp}}{(2\pi)^3}
E^{D/N}(z,\xi,\vec{T}_{\perp},\vec{\Delta}_{\perp})=
F_2^D(\Delta^2),\nonumber\\
\label{sum rule} \ea
where $F^D_i$ denotes diquark current contributions to the
nucleon form factor which are given in Appendix E.

\newpage
\section{Vertex corrections to the photon vertex}

The photon vertex correction, as shown in Fig. 2, consists
of the sum of a series of ring diagrams. Each diagram on
the left side of the diagrams in Fig. 2 is calculated as
follows:
\be \tau^a \gamma^{\mu}+
(-2)G_a\Pi_V^{\mu\nu}(\Delta)_{ab}\gamma_{\nu}\tau^b +
(-2)^2 G_a^2 \Pi_V^{\mu\nu}(\Delta)_{ab}
\Pi_{V\nu}^{\mu'}(\Delta)_{bc} \gamma_{\mu'}\tau^c
+\cdots, \ee
where
\be \Pi_V^{\mu\nu}(\Delta)_{ab}= 6i\delta_{ab}\int
\frac{d^4 k}{(2\pi)^4}tr_D[
S(k)\gamma^{\nu}S(k+\Delta)\gamma^{\mu}]. \ee
$\Pi_V^{\mu\nu}(\Delta)_{ab}$ can be decomposed into the
longitudinal and transverse parts:

\be\Pi_V^{\mu\nu}(\Delta)_{ab} \equiv \delta_{ab}
[(\Delta^2 g^{\mu\nu}-\Delta^{\mu}\Delta^{\nu})
\Pi_{V,T}(\Delta^2)+\Delta^{\mu}\Delta^{\nu}
\Pi_{V,L}(\Delta^2)].\ee
With the use of Ward identity, then it is clear that the
transverse part of $\Pi_V^{\mu\nu}(\Delta)_{ab}$ does not
contribute due to current conservation. Therefore the
series can be easily summed:
\be \rightarrow \frac{\tau^a \gamma^{\mu}}{1+2G_a \Delta^2
\Pi_{V,T}(\Delta^2)}, \label{vertex corrections} \ee
where $a=0 (i)$  means the isoscalar (isovector)
part and $G_{\omega,\rho}$ express the coupling constants
in the vector meson channels.

$\Pi_{V,L}(\Delta^2)$ can be calculated from
\be [(\Delta^2 g^{\mu\nu}-\Delta^{\mu}\Delta^{\nu})
\Pi_{V,T}(\Delta^2)+\Delta^{\mu}\Delta^{\nu}
\Pi_{V,L}(\Delta^2)]= 6i \int \frac{d^4 k}{(2\pi)^4}tr_D[
S(k)\gamma^{\nu}S(k+\Delta)\gamma^{\mu}].\ee
Inserting the PV-regularization factor, and performing the
$k$-integrals, we obtain
\ba \Pi_{V,T}(\Delta^2) = \frac{3}{\pi^2}\int_0^1 d\alpha
\alpha (1-\alpha) \left[
\frac{\Lambda^2}{M_Q^2-\alpha(1-\alpha)\Delta^2+\Lambda^2}
-ln \left(1+\frac{\Lambda^2}{M_Q^2-\alpha(1-\alpha)\Delta^2}
\right)
\right].\nonumber\\
\label{Pi_V} \ea
From Eqs. (\ref{vertex corrections}) and (\ref{Pi_V}), we can
easily see that there is no vertex correction at
$\Delta^2=0$, i.e., when the photon is on the mass shell.
\newpage
\section{Nucleon form factors}

Nucleon electromagnetic form factors can be calculated in
the same way as the GPDs, with the operator $\gamma^+ (1\pm
\tau_z)/2$ replaced by $\gamma^\mu(1\pm \tau_z)/2$. We
introduce Feynmann parameters $z,x_{1,2}$ to combine the
denominators of the propagators, and then a Wick rotation
is performed to obtain an Euclidean integral. The resulting
expressions are given by
\ba F_{1}^C (\Delta^2) &=&-4Q_q G_s Z_N \sum_i c_i
\int_{-\frac12}^{\frac12}dz \frac12
\int_0^{\infty}\frac{tdt}{8\pi^2}
\frac{\frac{t}{2}+(1-4z^2)\frac{\Delta^2}{4}+M_{Qi}^2}
{[t+M_{Qi}^2-(1-4z^2)\frac{\Delta^2}{4}]^2}\nonumber\\
F_{2}^C (\Delta^2)&=& -4Q_q G_s Z_N \sum_i c_i
\int_{-\frac12}^{\frac12}dz \frac12
\int_0^{\infty}\frac{tdt}{8\pi^2}
\frac{2M_Q M_N}{[t+M_{Qi}^2-(1-4z^2)\frac{\Delta^2}{4}]^2}
\nonumber\\
F_{1}^Q(\Delta^2)&=& Q_q g_d^2 Z_N \sum_i c_i\int_0^1 dx_1
\int_{-x_1}^{x_1}dx_2
\frac12 \int_0^{\infty}\frac{tdt}{8\pi^2}\nonumber\\
&\times&
 \frac{(1-x_1)^2 M_N^2+2(1-x_1)M_Q M_N+M_{Qi}^2+\frac{t}{2}
    +(x_1^2-x_2^2)\frac{\Delta^2}{4}}
{[t+(1-x_1)M_{Di}^2+x_1 M_{Qi}^2-x_1(1-x_1)M_N^2
-(x_1^2-x_2^2)\frac{\Delta^2}{4}]^3}\nonumber\\
F_{2}^Q(\Delta^2)&=& Q_q g_d^2 Z_N \sum_i c_i \int_0^1 dx_1
\int_{-x_1}^{x_1}dx_2
\frac12 \int_0^{\infty}\frac{tdt}{8\pi^2}\nonumber\\
&\times&
 \frac{2M_N x_1 [M_N(1-x_1)+M_Q]}
{[t+(1-x_1)M_{Di}^2+x_1 M_{Qi}^2-x_1(1-x_1)M_N^2
-(x_1^2-x_2^2)\frac{\Delta^2}{4}]^3}\nonumber\\
F_{1}^D(\Delta^2)&=& Q_d g_d^2 Z_N \sum_i c_i \int_0^1 dx_1
\int_{-x_1}^{x_1}dx_2
\frac12 \int_0^{\infty}\frac{tdt}{8\pi^2}\nonumber\\
&\times&
 \frac{2M_N (1-x_1)(x_1 M_N+M_Q)+\frac{t}{2}}
{[t+(1-x_1) M_{Qi}^2+x_1 M_{Di}^2 -x_1(1-x_1)M_N^2
-(x_1^2-x_2^2)\frac{\Delta^2}{4}]^3}\nonumber\\
F_{2}^D (\Delta^2)&=& Q_d g_d^2 Z_N \sum_i c_i \int_0^1 dx_1
\int_{-x_1}^{x_1}dx_2
\frac12 \int_0^{\infty}\frac{tdt}{8\pi^2}\nonumber\\
&\times&
 \frac{-2M_N (1-x_1) [x_1 M_N +M_Q]}
{[t+(1-x_1)M_{Qi}^2 +x_1 M_{Di}^2 -x_1(1-x_1)M_N^2
-(x_1^2-x_2^2)\frac{\Delta^2}{4}]^3},\nonumber\\
\ea
where C,P,D mean current, pole and diquark current contributions.

\end{document}